\documentclass[useAMS,usenatbib,usegraphicx]{mn2e}
\usepackage{bm}
\usepackage{times}

\begin{document}
\title[Far-infrared colours of blue compact dwarf galaxies]
{Effects of dust abundance on the far-infrared colours of blue
compact dwarf galaxies}
\author[H. Hirashita \& T. T. Ichikawa]{Hiroyuki Hirashita$^{1}$\thanks{E-mail:
    hirashita@asiaa.sinica.edu.tw} and Tomohiro T. Ichikawa$^{2}$
\\
$^1$ Institute of Astronomy and Astrophysics,
 Academia Sinica, P.O. Box 23-141, Taipei 10617, Taiwan \\
$^2$ Centre for Computational Sciences, University of Tsukuba,
     Tsukuba 305-8577, Japan
}
\date{2009 March 3}
\pubyear{2009} \volume{000} \pagerange{1}
\twocolumn

\maketitle \label{firstpage}
\begin{abstract}
We investigate the FIR properties of a sample of BCDs
observed by \textit{AKARI}. By utilizing the data at
wavelengths of $\lambda =65~\mu$m, 90 $\mu$m, and
140 $\mu$m, we find that the FIR colours of the BCDs are
located at the natural high-temperature extension of
those of the Milky Way and the Magellanic Clouds. This
implies that the optical properties of dust in BCDs are
similar to those in the Milky Way. Indeed, we explain
the FIR colours by assuming the same grain optical
properties, which may be appropriate for amorphous
dust grains, and the same size distribution as those
adopted for the Milky Way dust. Since both interstellar
radiation field and dust optical depth affect the dust
temperature, it is difficult to distinguish which of
these two physical properties is
responsible for the change of
FIR colours. Then, in
order to examine if the dust optical depth plays an important
role in determining the dust temperature, we investigate
the correlation between FIR colour (dust temperature) and
dust-to-gas ratio. We find that the dust temperature
tends to be high as the dust-to-gas ratio decreases but
that this trend cannot be explained by the effect of
dust optical depth. Rather, it indicates a correlation between
dust-to-gas ratio and interstellar radiation field.
Although the metallicity may also play a role in this
correlation, we suggest that the dust optical depth
could regulate the star formation activities, which govern
the interstellar radiation field. We also mention the
importance of
submillimetre data in tracing the emission from highly
shielded low-temperature dust.
\end{abstract}

\section{Introduction}

The far-infrared (FIR) emission from galaxies is often used
to trace the star formation activities
\citep{kennicutt98,inoue00,iglesias04}. Dust grains are the
source of FIR emission, and the strong connection between
FIR luminosity and star formation rate can be explained if
the ultraviolet (UV) light from massive stars is the
dominant source of dust heating. The FIR spectral energy
distribution (SED) of dust grains reflects various
information on the grains themselves and on the sources of
grain heating \citep[e.g.][]{takagi03,takeuchi05,dopita05}.
Dust temperature is determined by the intensity of
interstellar radiation field (ISRF) and the optical
properties of dust
(i.e.\ how it absorbs and emits light). Since dust
grains absorb UV light efficiently, the FIR luminosity
and the dust temperature mostly reflect the UV radiation
field \citep{buat96}.

Observationally, dust temperature can be estimated from
FIR colour, which is defined as the flux ratio between two
FIR wavelengths. If more than two ($\geq 3$) bands are
available in FIR, we can take two independent FIR
colours and examine a FIR colour--colour relation. A
colour--colour relation,
because of increased information compared with a single
FIR colour, enables us to obtain not only the dust
temperature but also the wavelength dependence of dust
emissivity. \citet{nagata02}, adopting 60 $\mu$m,
100 $\mu$m, and 140 $\mu$m as three wavelengths, show
that there is a tight relation in the FIR colour--colour diagram.
This tightness implies a common wavelength dependence of
FIR emissivity among various galaxies. Their work has been
developed by \citet{hibi06}, who show that a tight FIR
colour--colour relation found for the Milky Way dust
emission, called ``main correlation'', is also consistent with
the FIR colour--colour relation of a sample of nearby
galaxies (see also \citealt{sakon04,onaka07}). As stated by
\citet{hibi06},
this implies that the optical properties of grains in
FIR are common
among the Milky Way and nearby galaxies.

Since the FIR colour--colour relation can be used to
constrain the optical properties of dust grains,
the FIR colour--colour relation of galaxies with
various evolutionary stages provides pieces of
information on the grain evolution in
galactic environments.
In particular, the dominant production
source of dust grain could be different in early
phase of galaxy evolution
\citep{todini01,nozawa03,maiolino04,schneider04}.
Moreover, the effects of interstellar
processing of dust grains, especially accretion of
heavy elements onto dust grains, coagulation, and
shattering, depend on metallicity (or dust abundance).
Thus, it is probable that the grain properties
evolve as galaxies evolve.

Although it is difficult to obtain the rest-frame FIR
data of distant galaxies in an early phase of galaxy
evolution, there are nearby possible ``templates'' of
primeval galaxies, blue compact dwarf galaxies (BCDs).
Indeed, BCDs have on-going star formation in metal-poor
and gas-rich environments \citep{sargent70,vanzee98},
which could be similar to those in high-redshift
star-forming galaxies. Moreover, BCDs harbour an
appreciable amount of dust \citep[e.g.][]{thuan99} and
can be used to
investigate the dust properties in chemically unevolved
galaxies \citep{takeuchi05}.

It is possible to investigate the metallicity dependence
of dust properties by sampling BCDs with a variety of
metallicity. Recently \citet{engelbracht08} have
examined the metallicity dependence of dust emission in
various wavelengths by using \textit{Spitzer} data. In
FIR, they have used Multiband Imaging Photometer for
\textit{Spitzer} (MIPS) 70 $\mu$m and 160 $\mu$m bands,
and have shown that there is a correlation between
dust temperature derived from these two bands and
metallicity. This indicates the importance of studies on
dust emission in a wide metallicity range.

The \textit{AKARI} satellite \citep{murakami07} provides
us with a good opportunity to study FIR colour--colour
relations, since Far-Infrared Surveyor (FIS) on
\textit{AKARI} has four bands in FIR (65 $\mu$m,
90 $\mu$m, 140 $\mu$m, and 160 $\mu$m)
\citep{murakami07,kawada07}. Indeed,
seven of the eight BCDs in
\citet[hereafter H08]{hirashita08} are detected at
65~$\mu$m, 90~$\mu$m, and 140~$\mu$m. This indicates
that it is possible to study the FIR colour--colour
relation of BCDs by using \textit{AKARI} data. In this
paper, we add four more BCDs available in the
\textit{AKARI} archive and examine the FIR colour--colour
relation of BCDs.
Then, we extract information on what determines or
regulates the FIR emission in metal-poor environments
with the aid of theoretical models for dust emission.

This paper is organized as follows. First, in
Section~\ref{sec:model}, we describe the models of
dust emission with a simple radiative transfer recipe.
Then, in Section~\ref{sec:data}, we explain
the data analysis of the BCD sample observed
by \textit{AKARI}. In
Section~\ref{sec:result}, we overview the
results of the model calculations in comparison with
the observational data.
In Section~\ref{sec:discussion}, we discuss our results,
focusing on the relation between FIR colour and dust
content. Finally, the conclusion is presented in
Section~\ref{sec:conclusion}.

\section{FIR SED Model}\label{sec:model}

We adopt the theoretical framework of \citet{draine01}
to calculate the SED of dust emission in FIR. The
framework has already been described in
\citet*[hereafter HHS07]{hirashita07}, but we modify
it to treat BCDs in this paper. Moreover,
 {as described in the last paragraph in
Sect.\ \ref{subsec:isrf},}
we newly include the effect of radiative transfer
in a similar way to \citet{galliano03}.
We overview our framework, focusing on the
change from HHS07.

\subsection{Interstellar radiation field}
\label{subsec:isrf}

The stellar SEDs of BCDs are generally harder than
those of spiral galaxies \citep*{boselli03}.
\citet{madden06} also show a hard stellar radiation
field from the observation of mid-infrared emission
lines in low-metallicity star-forming galaxies. Thus,
we change the SED of ISRF, and adopt the following
fitting formula according to the averaged SED of the
BCD sample in \citet{boselli03}:
\begin{eqnarray}
4\pi\lambda J_\lambda^{(0)} =\left\{
\begin{array}{l}
0 ~~~~~ (\lambda_{\mu\mathrm{m}}\leq 0.0912), \\
3.02\times 10^{-4}\lambda_{\mu\mathrm{m}}^{-1.20}\\
~~~~~~~(0.0912<\lambda_{\mu\mathrm{m}}\leq 0.25), \\
1.55\times 10^{-3}\times 10^{-3}
\lambda_{\mu\mathrm{m}}^{-0.0371}\\ ~~~~~~~
(0.25<\lambda_{\mu\mathrm{m}}\leq 0.37), \\
2.39\lambda_{\mu\mathrm{m}}^{7.41}~~~~~
(0.37<\lambda_{\mu\mathrm{m}}\leq 0.4), \\
4\pi\lambda [W_1B_\lambda (T_1)+W_2B_\lambda (T_2)]
\\ ~~~~~~~ 
(0.4<\lambda_{\mu\mathrm{m}}),
\end{array}
\right.\label{eq:isrf}
\end{eqnarray}
where $J_\lambda^{(0)}$ is the ISRF intensity whose
normalization is determined so that the intensity at
wavelength $\lambda =0.2~\mu$m (almost the centre of
the UV range; \citealt{buat96}) is equal to the solar
neighborhood ISRF estimated by \citet*{mathis83}
($4\pi\lambda J_\lambda^{(0)}$ is expressed in units
of erg cm$^{-2}$ s$^{-1}$),
$\lambda_{\mu\mathrm{m}}$ is the wavelength
in units of $\mu$m, $B_\lambda (T)$ is the Planck
function, $(T_1,\, T_2)=(1000,\, 4000~\mathrm{K})$, and
$(W_1,\, W_2)=(1.61\times 10^{-15},\, 2.90\times 10^{-14})$.

In this paper, we newly include the effect of dust
extinction on the ISRF. We simply assume that the ISRF
intensity, $J_\lambda$, at optical depth $\tau_\lambda$
is
\begin{eqnarray}
J_\lambda=J_\lambda^{(0)}\chi\exp(-\tau_\lambda)\, ,
\end{eqnarray}
where $\chi$ is the scaling factor of the ISRF relative to
$J_\lambda^{(0)}$ and is assumed to be independent of
wavelength. For the wavelength dependence of
$\tau_\lambda$, we assume the Milky Way extinction
curve taken from \citet*{cardelli89} with $R_V=3.1$. Thus,
the extinction at any wavelength $\lambda$, $A_\lambda$
(in units of magnitude), can be determined if we set a value
of $A_V$ (extinction at $V$ band). The optical depth can
be related with the extinction as
$\tau_\lambda =A_V/1.086$. The following results does not
change drastically if we assume an extinction curve
appropriate for the Large Magellanic Clouds (LMC) or the
Small Magellanic Cloud (SMC), since the detailed shape
of UV--optical SED is not important in this paper.
Strictly speaking, we
treat scattering as effective absorption, and this treatment
would overestimate the absorbed energy. Thus, we
interpret $A_V$ given here as effective
absorption optical depth, so that the total energy
absorbed by dust in our formulation is equal to the
energy that would be absorbed if scattering were properly
treated.

\citet{galliano03} also adopted a similar treatment, which
is valid if dust is
distributed in a thin shell surrounding the stars. Such a
``screen'' geometry
enhances the effect of dust extinction (i.e., radiative
transfer) in comparison with
a mixed geometry between stars and dust.
Thus, our models are suitable to examine the extent to
which the radiative transfer effects could affect the
FIR SEDs.

\subsection{Grain properties}\label{subsec:properties}

We consider silicate and graphite grains in this paper.
Since we are interested in the wavelength range
appropriate for the \textit{AKARI} FIS bands
($\lambda\sim 50$--180 $\mu$m;
Section \ref{subsec:analysis}), we neglect polycyclic
aromatic hydrocarbons (PAHs), which contribute
significantly to the emission at
$\lambda\la 20~\mu$m \citep{desert90,dwek97,draine01}.

We assume a grain to be spherical with a radius of $a$.
The absorption cross section of the grain is expressed
as $\pi a^2Q_\mathrm{abs}(\lambda )$, where
$Q_\mathrm{abs}(\lambda )$ is called absorption
efficiency. We adopt the optical constants of
astronomical silicate and graphite for
$\lambda <100~\mu$m from \citet{draine84} and
\citet{weingartner01} and calculate the absorption
efficiency by using Mie theory \citep{bohren83}.
For the absorption efficiency at $\lambda >100~\mu$m,
we adopt a functional form proposed by
\citet{reach95}:
\begin{eqnarray}
Q_\mathrm{abs}(\lambda )=
\frac{(\lambda /\lambda_0)^{-2}}
{[1+(\lambda_1/\lambda)^6]^{1/6}}\, ,\label{eq:reach}
\end{eqnarray}
which behaves like $\beta =1$
($Q_\mathrm{abs}(\lambda )\propto\lambda^{-\beta}$,
where $\beta$ is called emissivity index) for
$\lambda\ll\lambda_1$ and $\beta =2$ at
$\lambda\gg\lambda_1$ \citep[see also][]{bianchi99}.
We assume that $\lambda_1=200~\mu$m, following
\citet{reach95}. We set the value of $\lambda_0$ so that
the continuity of $Q_\mathrm{abs}$ at $100~\mu$m is
satisfied. As shown in HHS07, the wavelength dependence of
$Q_\mathrm{abs}$ assumed here explains the FIR
colour--colour
relations of the Milky Way, the LMC and the SMC.
Although the absorption efficiency at $\lambda >100~\mu$m
is modified, we call the grain species ``silicate'' and
``graphite'' according to the adopted absorption
efficiency at $\lambda <100~\mu$m.

The number density of grains with sizes between $a$ and
$a+\mathrm{d}a$ is denoted as $n_i(a)\,\mathrm{d}a$, where
the subscript $i$ denotes a grain species (silicate or
graphite). Since there is little knowledge on the grain
size distribution in BCDs, we simply assume a power-law
form for $n_i(a)$:
\begin{eqnarray}
n_i(a)=\mathcal{C}_ia^{-K}~~~(a_\mathrm{min}\leq a\leq
a_\mathrm{max})\, ,
\label{eq:size_dist}
\end{eqnarray}
where $a_\mathrm{min}$ and $a_\mathrm{max}$ are the upper
and lower cutoffs of grain size, respectively, and
$\mathcal{C}_i$ is the normalizing constant. We assume
$K=3.5$, $a_\mathrm{min}=3.5$ \AA\ and
$a_\mathrm{max}=0.25~\mu$m for both
graphite and silicate \citep*{mathis77,li01}.
As long as we treat the FIR colours of a single species
(Section \ref{subsec:colour}), the constant
$\mathcal{C}_i$ cancels out. Because the dust
composition in BCDs is still unclear, we consider
silicate and graphite separately as two possible dust
species.

Following HHS07, we
adopt the multi-dimensional Debye models for the heat
capacities of grains \citep{draine01}. The physical
parameters necessary to characterize the heat capacities
can be found in HHS07.

\subsection{Calculation of FIR colours}
\label{subsec:colour}

The FIR intensity $\mathcal{J}_\nu^i(\lambda )$ (per
frequency per hydrogen nucleus per solid angle) of the
emission from dust species $i$ at
a wavelength $\lambda$ can be estimated as
\begin{eqnarray}
\mathcal{J}_\nu^i(\lambda )=
\int_{a_\mathrm{min}}^{a_\mathrm{max}}\mathrm{d}
a\frac{1}{n_\mathrm{H}}n_i(a)\pi a^2Q_\mathrm{abs}
(\lambda )\int_0^\infty\mathrm{d}T\, B_\nu (T)
\frac{\mathrm{d}P_i}{\mathrm{d}T},\hspace{-1cm}\nonumber\\
\end{eqnarray}
where $B_\nu (T)$ is the Planck function at frequency $\nu$
and temperature $T$, and $\mathrm{d}P_i/\mathrm{d}T$ is the
temperature distribution function of the grains under the
ISRF evaluated in Section \ref{subsec:isrf}. The
calculation method of
$\mathrm{d}P_i/\mathrm{d}T$ is based on the concept of
stochastic heating, and we follow
the formalism of \citet{draine01}
(see their section 4).
We assume that the FIR radiation
is optically thin. Thus, after the radiative transfer effect,
the intensity at $\lambda$ per frequency per hydrogen nucleus
per solid angle, $I_\nu (\lambda )$, becomes
\begin{eqnarray}
I_\nu (\lambda)=\int_0^{s}\mathcal{J}_\nu\,\mathrm{d}s\, ,
\end{eqnarray}
where $s$ is the path length, which is proportional to
$A_V$. Here the colour at wavelengths of $\lambda_1$ and
$\lambda_2$ is defined as
$I_\nu (\lambda_1)/I_\nu (\lambda_2)$, which is called
$\lambda_1-\lambda_2$ colour and denoted as
$(\lambda_1/\lambda_2)_\mathrm{cl}$ (we omit $\mu$m in
this expression; e.g.\ $(60/100)_\mathrm{cl}$ and
$(140/100)_\mathrm{cl}$). Since we only consider colours
(i.e.\ flux ratios), the proportionality
constant between $s$ and $A_V$ does not matter in this
paper. Thus, practically, we can replace $s$ with $A_V$.

 {
Since we calculate colours in this paper, we can define
the colour with $A_V=0$ as the optically thin extreme
($\lim_{A_V\to 0}(\lambda_1/\lambda_2)_\mathrm{cl}$).
In other words, the FIR colour calculated with $A_V=0$
(in fact $A_V\to 0$) represents the case where the dust
is illuminated by a radiation field without extinction.
In the following, $A_V=0$ means that the dust is illuminated
by a radiation field without extinction, not that there is
no dust.
}

\section{Data}\label{sec:data}

\subsection{Analysis of the \textit{AKARI} data}
\label{subsec:analysis}

H08 analysed eight BCDs observed by FIS onboard
\textit{AKARI} with four photometric bands of \textit{N60},
\textit{WIDE-S}, \textit{WIDE-L}, and \textit{N160},
whose central wavelengths are 65, 90, 140, and
160 $\mu$m with effective band widths of
$\Delta\lambda =21.7$, 37.9, 52.4, and 35.1 $\mu$m,
respectively \citep{kawada07}. The measured FWHMs of
the point spread function are $37\pm 1''$, $39\pm 1''$,
$58\pm 3''$, and
$61\pm 4''$, respectively, for the above four bands.
We exclude Mrk 36 from the sample, since it is not
clearly detected at \textit{WIDE-L} (140 $\mu$m).
The other seven sample BCDs in H08 are listed in the
lower part of Table \ref{tab:flux}.

\begin{table*}
\centering
\begin{minipage}{110mm}
\caption{Measured fluxes (``$<$'' indicates the 3 $\sigma$ upper
limit). The upper four objects are newly analysed in this paper,
while the lower seven are the sample in \citet{hirashita08}.}
\label{tab:flux}
    \begin{tabular}{lcccc}
      \hline
      Name     & \textit{N60} & \textit{WIDE-S}  &
               \textit{WIDE-L} & \textit{N160}
               \\
               & ($\lambda =65~\mu$m) & ($\lambda =90~\mu$m) &
               ($\lambda =140~\mu$m) & ($\lambda =160~\mu$m)
               \\ \hline 
      UM 420   & $0.46\pm 0.05$ Jy & $0.37\pm 0.04$ Jy &
               $0.14\pm 0.03$ Jy & $<0.36$ Jy \\
      Mrk 59   & $2.3\pm 0.2$ Jy & $2.1\pm 0.2$ Jy &
               $1.4\pm 0.3$ Jy & $1.2\pm 0.3$ Jy \\
      Mrk 487  & $0.45\pm 0.05$ Jy & $0.43\pm 0.04$ Jy &
               $0.21\pm 0.06$ Jy & $<0.69$ Jy \\
      SBS 1319+579 & $0.38\pm 0.04$ Jy & $0.29\pm 0.03$ Jy &
               $0.28\pm 0.06$ Jy & $<0.59$ Jy \\ \hline
      II Zw 40 & $6.9\pm 0.7$ Jy & $6.6\pm 0.7$ Jy &
               $3.7\pm 0.7$ Jy & $3.4\pm 0.9$ Jy \\
      Mrk 7    & $0.83\pm 0.23$ Jy & $0.84\pm 0.08$ Jy &
               $1.1\pm 0.2$ Jy & $1.0\pm 0.3$ Jy \\
      Mrk 71   & $3.7\pm 0.4$ Jy & $3.1\pm 0.3$ Jy &
               $2.2\pm 0.4$ Jy & $<3.0$ Jy \\
      UM 439   & $0.42\pm 0.04$ Jy & $0.40\pm 0.04$ Jy &
               $0.31\pm 0.06$ Jy & $<0.45$ Jy \\
      UM 533   & $0.54\pm 0.05$ Jy & $0.61\pm 0.06$ Jy &
               $0.54\pm 0.11$ Jy & $<0.68$ Jy \\
      II Zw 70 & $0.89\pm 0.09$ Jy & $0.80\pm 0.24$ Jy &
               $0.75\pm 0.15$ Jy & $0.62\pm 0.16$ Jy \\
      II Zw 71 & $0.32\pm 0.03$ Jy & $0.54\pm 0.05$ Jy&
               $1.3\pm 0.26$ Jy & $1.3\pm 0.3$ Jy \\
      \hline
    \end{tabular}
\end{minipage}
\end{table*}

In this paper, we add four BCDs available in the
\textit{AKARI}
archive\footnote{http://darts.isas.jaxa.jp/astro/akari/.}:
UM 420, Mrk 59, Mrk 487, and SBS 1319+579 as shown
in the upper part of Table \ref{tab:flux}. We selected
these galaxies among the BCDs observed by
FIS after investigating if they were
detected at \textit{N60} (65 $\mu$m), \textit{WIDE-S}
(90 $\mu$m), and \textit{WIDE-L} (140 $\mu$m).
The observing mode (FIS01), the scan speed
(8$''$ s$^{-1}$), and the reset interval (2 s) are
the same as those adopted for the sample in H08.
The method of data reduction and analysis is the
same as that in H08, and the measured fluxes are listed
in Table \ref{tab:flux}. Below we briefly review the data
reduction and analysis (see H08 for details).

The raw data were reduced by using the FIS Slow Scan Tool
(version
20070914).\footnote{http://www.ir.isas.ac.jp/ASTRO-F/Observation/.}
Because the detector response is largely affected by a hit
of high energy ionizing particle, we used the local flat;
that is, we corrected the detector sensitivities by assuming
uniformity of the sky brightness
\citep*{verdugo07}.\footnote{We referred to \textit{AKARI} FIS
Data User Manual Version 2 for the data reduction and
analysis (http://www.ir.isas.ac.jp/ASTRO-F/Observation/). When
we ran the pipeline command {\tt ss\_run\_ss}, we applied
options {\tt /local,/smooth,width\_filter=80}: The flat field was
built from the observed sky, and boxcar smoothing with a filter
width of 80 s in the time series data was applied to remove
remaining background offsets among the pixels.}
The errors caused by the smoothing procedures are well
within 10\% for \textit{N60} ($65~\mu$m) and
\textit{WIDE-S} ($90~\mu$m), 20\% for
\textit{WIDE-L} ($140~\mu$m), and 25\% for
\textit{N160} ($160~\mu$m). These error levels are
consistent to or
somewhat larger than the background fluctuation
mainly caused by a hit of high energy ionizing particle,
which implies that the background fluctuation is a major
component in the errors. The above values are adopted for
the errors, but we adopt larger errors for some galaxies
which have larger background noises.
At \textit{N160} (160~$\mu$m), some objects are not
detected. For them,
we adopt 3 times the background uncertainty as an upper
limit. Since the sample BCDs are compact enough to be treated
as point sources with the resolution of FIS, we follow the
point-source photometry described in \citet{verdugo07}.

Finally  {the same colour correction as in H08 was
adopted} for the \textit{WIDE-L} ($140~\mu$m) fluxes: the
colour correction factor was assumed to be 0.93 (the flux was
divided by this factor). We did not apply colour correction to
the \textit{N60} (65 $\mu$m), \textit{N160} (160 $\mu$m), and
\textit{WIDE-S} ($90~\mu$m) fluxes, following H08.
For these three bands, the uncertainty caused by not
applying colour correction is smaller than
the errors put in Table \ref{tab:flux}.

\subsection{FIR colours}

The FIR colours are defined by the flux ratio between two
bands (Section \ref{subsec:colour}). HHS07 investigated the
FIR colours of the Milky Way and the Magellanic Clouds by
using the Zodi-Subtracted Mission Average (ZSMA) taken by
the Diffuse Infrared Background Experiment (DIRBE) of the
\textit{Cosmic Background Explorer} (\textit{COBE}).
Details of the observational data analysis can be found
in \citet{hibi06}, who adopted DIRBE bands of
60~$\mu$m, 100~$\mu$m, and 140~$\mu$m and
two colours, 60~$\mu$m--100~$\mu$m colour,
$(60/100)_\mathrm{cl}$, and 140~$\mu$m--100~$\mu$m
colour, $(140/100)_\mathrm{cl}$.

\begin{figure}
\begin{center}
\includegraphics[width=8.5cm]{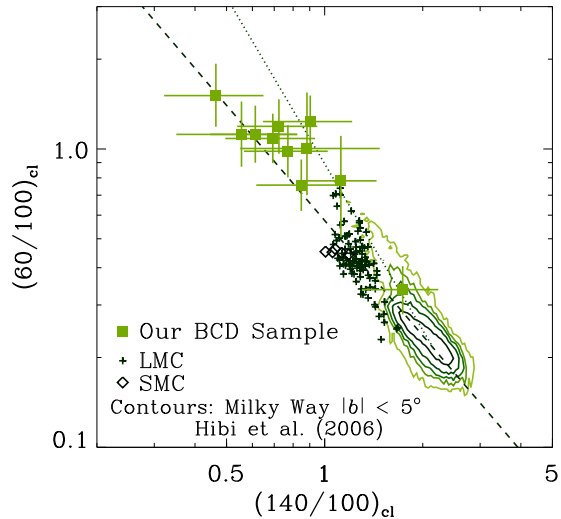}
\end{center}
\caption{Far-infrared colour--colour diagram showing the
relation between 60 $\mu$m--100 $\mu$m colour and
140 $\mu$m--100 $\mu$m colour. The filled squares with
error bars indicate the current BCD sample, while the
crosses and open diamonds show the DIRBE data of the LMC
and the SMC, respectively, taken from \citet{hibi06}.
We also present the Milky Way data (with Galactic latitudes
of $|b|<5^\circ$) adopted from \citet{hibi06} by
the contours, which show the levels where 50, 80, 90,
95, and 99 per cent of the data are contained. The dashed
line and the dotted lines present the fittings for the main
correlation (equation \ref{eq:main}) and the subcorrelation
(equation \ref{eq:sub}), respectively.
\label{fig:clrdat}}
\end{figure}

We can also take two colours for our BCD sample. However,
the wavelengths of the FIS bands are slightly different
from the DIRBE bands. Thus, we apply the following
corrections. First, we fit $A\nu^\beta B_\nu (T_\mathrm{d})$
($A$ is a constant and
$\beta$ is the emissivity index) to the 65 $\mu$m and
90 $\mu$m fluxes derive the dust temperature
$T_\mathrm{d}$. We denote the dust temperature obtained
in this way as $T_\mathrm{d}((65/90)_\mathrm{cl},\,\beta )$.
We adopt $\beta =1$ and 2. Then, we estimate the
60 $\mu$m flux by evaluating
$A\nu^\beta B_\nu (T_\mathrm{d})$ at
$\lambda =60~\mu$m under the values of $A$ and
$T_\mathrm{d}((65/90)_\mathrm{cl},\,\beta )$ obtained
above. The same procedure is applied for the
140 $\mu$m and 90 $\mu$m fluxes to obtain the
100 $\mu$m flux. In this way, we can obtain
$(60/100)_\mathrm{cl}$ and
$(140/100)_\mathrm{cl}$
for the BCD sample. In Fig.\ \ref{fig:clrdat}, we show
the colour--colour
relation for the sample.

For comparison, we also show the data of \citet{hibi06}
for the Milky Way (the Galactic plane with Galactic
latitudes of $|b|<5^\circ$) and the Magellanic Clouds in
Fig.\ \ref{fig:clrdat}. \citet{hibi06} found that more than
90\% of the data lie on a strong correlation called main
correlation, which can be fitted as
\begin{eqnarray}
(140/100)_\mathrm{cl}=0.65(60/100)_\mathrm{cl}^{-0.78}
\, .\label{eq:main}
\end{eqnarray}
The main correlation also explains the FIR colours of
Galactic high latitudes ($|b|>5^\circ$), the LMC, and
the SMC (\citealt{hibi06,hibiphd06}; HHS07).
In the Galactic plane, there is another correlation
sequence, called subcorrelation in \citet{hibi06}:
\begin{eqnarray}
(140/100)_\mathrm{cl}=0.93(60/100)_\mathrm{cl}^{-0.56}
\, .\label{eq:sub}
\end{eqnarray}
The subcorrelation is not seen in the high Galactic
latitudes (\citealt{hibiphd06}; HHS07). This supports
the idea of \citet{hibi06} that the
subcorrelation is produced by a contamination of
high-ISRF regions, which tend to reside in the Galactic
plane.

It is interesting that not only the LMC and the SMC but
also the current BCD sample has consistent FIR
colours to the main correlation or the subcorrelation
(Fig.\ \ref{fig:clrdat}). This implies that the
wavelength dependence of the FIR
emissivity is not different among the BCDs, the LMC,
the SMC, and the Milky Way.
In the following
section, we examine if the FIR colours of BCDs can
really be reproduced with the emission properties
adopted by HHS07, who explained the FIR colours of
the Milky Way, the LMC, and the SMC
(Section \ref{sec:model}).

\section{Results}\label{sec:result}

\subsection{Interstellar radiation field}
\label{subsec:dep_isrf}

\begin{figure}
\begin{center}
\includegraphics[width=8.5cm]{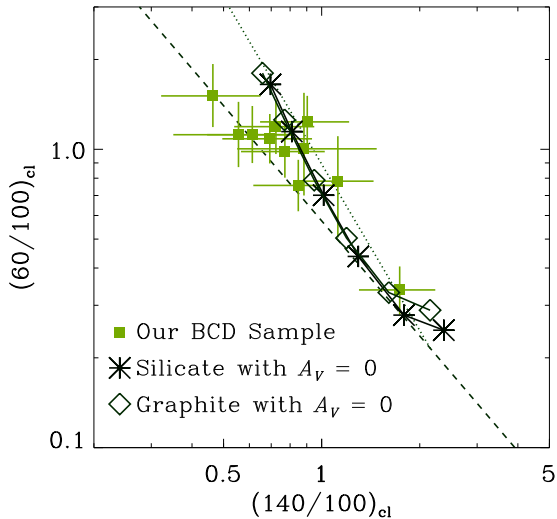}
\end{center}
\caption{Far-infrared colour--colour relations for silicate
and graphite (asterisks and open diamonds, respectively).
The different points correspond to $\chi =1$, 3, 10, 30,
100, and 300 from high to low [140/100].
 {We adopt $A_V=0$ (optically thin extreme) for
the ISRF.}
The same data as
shown in Fig.~\ref{fig:clrdat} as well the the main
correlation and subcorrelation (dashed and dotted lines,
respectively) are also presented.
\label{fig:clr_Av0}}
\end{figure}

We present the dependence of FIR colours on ISRF.
Here we adopt $A_V=0$  {(i.e.\ the optically thin
extreme for the ISRF; Section \ref{subsec:colour})} to
concentrate only on the effects of ISRF intensity.
In Fig.\ \ref{fig:clr_Av0}, we show the FIR
colour--colour relation ($(60/100)_\mathrm{cl}$ and
$(140/100)_\mathrm{cl}$) calculated for silicate and
graphite. We observe that the
FIR colours of the BCD sample can be reproduced with
$\chi\sim 30$--300 except for the lowest data point
(II Zw 71). This is consistent with the conclusion in H08
that the ISRF (especially the UV ISRF,
which contributes most to the dust heating) in BCDs can
be 100 times higher than the Galactic ISRF in the solar
neighbourhood.

Our results not only confirm the warm \textit{IRAS}
(\textit{Infrared Astronomical Satellite})
60 $\mu$m--100 $\mu$m colours of BCDs
\citep{hoffman89}, but also show high temperatures of
dust contributing to
the emission at $\lambda >100~\mu$m. A few
Virgo Cluster BCDs observed by \textit{ISO}
(\textit{Infrared Space Observatory}) show relatively
cold 170 $\mu$m--100 $\mu$m colours \citep{popescu02},
indicating that the presence of warm dust is not
necessarily the case at least in cluster environments.

We also find that the calculated FIR colours cover the
area between the main correlation and the subcorrelation.
With $\chi\la 10$, the colours trace the main correlation,
which is appropriate for the Milky Way, the LMC and
the SMC. On the other hand, the FIR colours are rather
similar to the subcorrelation for $\chi\ga 100$.
While \citet{hibi06} and HHS07 claimed that the
subcorrelation can be reproduced with dust emission
with multiple temperature components, we see here that
the upper part of the subcorrelation can be explained with
a single temperature.

\subsection{Radiative transfer effects}\label{subsec:radtr}

 {If radiative transfer effects are considered,} FIR
emission comes from dust at different
optical depths, where the ISRF intensities are different
because of dust extinction. As modeled in
Section \ref{subsec:colour}, the FIR intensity is the
sum of all contributions from different optical
depths. \citet{hibi06} expect that radiation transfer
effects shift the FIR colours toward the subcorrelation on
the colour--colour diagram.

\begin{figure*}
\begin{center}
\includegraphics[width=8.5cm]{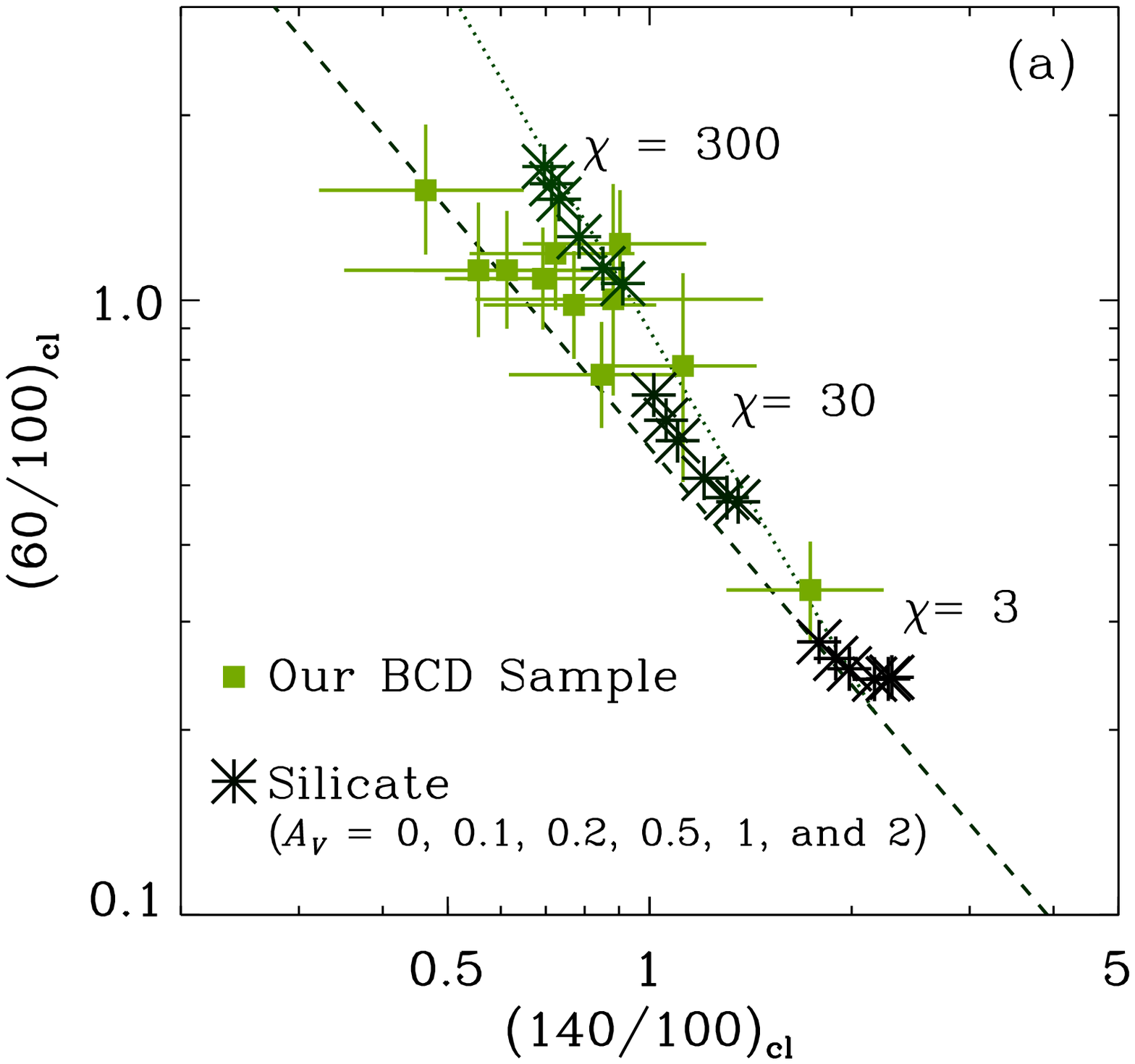}
\includegraphics[width=8.5cm]{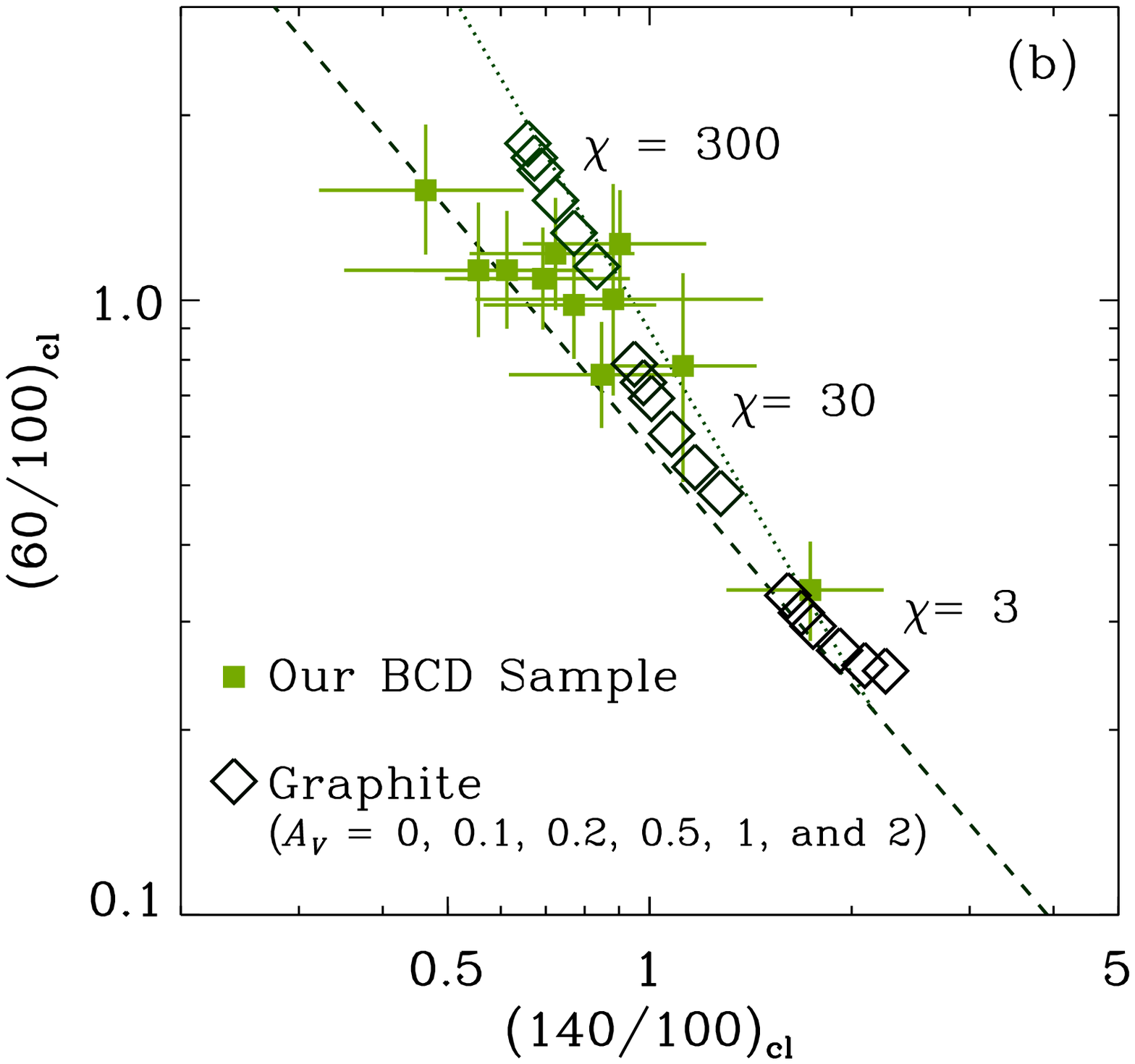}
\end{center}
\caption{Far-infrared colour--colour relations for (a)
silicate and (b) graphite (asterisks and open diamonds,
respectively). The results for $\chi =3$, 30, and 300 are
shown. For each value of $\chi$,
$A_V=0$, 0.1, 0.2, 0.5, 1, and 2 correspond
to the data points from low to high
$(140/100)_\mathrm{cl}$.  {Note that $A_V=0$ is
the optically thin extreme for the ISRF}. The same
data as shown in Fig.~\ref{fig:clrdat} as well the the main
correlation and subcorrelation (dashed and dotted lines,
respectively) are also presented.
\label{fig:clr_varAv}}
\end{figure*}

In Fig.\ \ref{fig:clr_varAv}, we show the FIR colours
for $\chi =3$, 30, and 300 with various values of
$A_V$ ($A_V=0$, 0.1, 0.2, 0.5, 1, and
2)\footnote{ {As stressed in Section \ref{subsec:colour},
the case of $A_V=0$ is considered as the optically thin
extreme for the ISRF.}}. The FIR
colours change little for $A_V\ga 2$, since the
emission is negligible from such large optical
depths in the wavelength range of interest because of
low dust temperatures. As expected by \citet{hibi06},
there is a slight trend that the colours move toward the
subcorrelation for $\chi\la 30$. However,
we observe that the radiative
transfer effects shift the FIR colours along the main
correlation or the subcorrelation for $\chi\ga 30$.
This indicates that the radiative
transfer effects are hard to be distinguished from the
change of $\chi$ on the colour--colour diagram.

How can we distinguish the variation of $A_V$ from
that of $\chi$? We expect that if the change of $A_V$
controls the colour variation there may be a
correlation between dust content and FIR colours.
Below we further discuss the observed FIR colours in
BCDs in terms of the dust content.

\section{Discussion}\label{sec:discussion}

\subsection{Dust content and FIR colours}
\label{subsec:dust_content}

The dust mass of the current BCD sample is estimated by
the \textit{WIDE-S} ($90~\mu$m) and \textit{WIDE-L}
($140~\mu$m) fluxes, which are suitable to trace the
total mass of large grains which have the dominant
contribution to the total dust content
\citep{galliano05}.\footnote{\citet{galliano05} also
introduce a very cold dust component contributing to the
submillimetre flux, but its abundance is
sensitive to the assumed emissivity index of large
grains (Section \ref{subsec:submm}). Thus, we here
concentrate on the dust
component traced in FIR and do not consider very cold
dust.} The dust mass $M_\mathrm{d}$ is related to the
flux as (e.g.\ \citealt{hildebrand83}; H08)
\begin{eqnarray}
M_\mathrm{d}=
\frac{F_\nu (\lambda )D^2}{\kappa_\nu B_\nu (T_\mathrm{d})}
\, ,\label{eq:mdust}
\end{eqnarray}
where $\kappa_\nu$ is the mass absorption coefficient of
dust grains. The mass absorption coefficient is related
with the absorption efficiency $Q_\nu$ as (e.g.\ H08)
\begin{eqnarray}
\kappa_\nu =\frac{3Q_\nu}{4as}\, ,
\end{eqnarray}
where $s$ is the grain material density. If the grain
radius is much smaller than the wavelength, $Q_\nu /a$
is independent of $a$ \citep{hildebrand83}.
The values adopted in
Section \ref{subsec:properties} indicate
$Q_\nu /a=1.14\times 10^{-2}~\mu\mathrm{m}^{-1}$ for
silicate and $1.23\times 10^{-2}~\mu\mathrm{m}^{-1}$
for graphite at $\lambda =125~\mu$m. By adopting
typical densities for silicate and graphite as
$s=3.3$ g cm$^{-3}$ and
2.25 g cm$^{-3}$, respectively, we obtain the
mass absorption coefficients at $\lambda =125~\mu$m
as $\kappa_{125}=25.9$ and 41.0 cm$^2$ g$^{-1}$ for
silicate and graphite, respectively.
H08 derived $\kappa_{125}=18.8$ cm$^2$ g$^{-1}$ based
on \citet{hildebrand83}.

Following H08, we adopt the following expression for
$\kappa_\nu$ to estimate the dust mass from the
\textit{AKARI} data:
\begin{eqnarray}
\kappa_\nu =\kappa_{125}\left(
\frac{\lambda}{125~\mu\mathrm{m}}\right)^{-\beta}\, .
\end{eqnarray}
The absorption coefficient adopted in
Section \ref{subsec:properties} cannot be fitted with a
single power law, but the index lies between
$\beta =1$ and 2. Thus, if we adopt the wavelength
dependence described in Section \ref{subsec:properties},
the estimated dust mass should be between the dust
masses obtained with $\beta =1$ and $\beta =2$. We prefer
to adopt this simple power-law expression for $\kappa_\nu$
here, since it can be directly compared with the results
in H08 and the simple expression is useful for
observational usage.

In Table \ref{tab:mdust}, we list the dust mass estimated
from equation (\ref{eq:mdust}). We determine the dust
temperature $T_{\rm d}$ by fitting
$A\nu^{\beta}B_\nu (T_{\rm d})$ ($A$ is a constant) to the
90 $\mu$m and 140 $\mu$m fluxes. Note that $A$ can be
eliminated if we take the flux ratio of those two bands. The
estimated dust temperatures are listed in
Table \ref{tab:mdust}, where $T_\mathrm{d}(\beta =1)$
and $T_\mathrm{d}(\beta =2)$
denote the dust temperatures evaluated with
$\beta =1$ and 2, respectively. For the mass absorption
coefficient, we adopt $\kappa_{125}=25.9$ cm$^2$ g$^{-1}$
(i.e.\ the value for silicate), but the dust mass is simply
proportional to $\kappa_{125}^{-1}$ if another
value of $\kappa_{125}$ is adopted. The distance is taken
from H08 for the common sample, and is estimated by
using the Galactocentric velocity taken from the NASA/IPAC
Extragalactic Database
(NED)\footnote{http://nedwww.ipac.caltech.edu/.}
by assuming a Hubble constant of 75 km s$^{-1}$ Mpc$^{-1}$.
The dust masses estimated with $\beta =1$ and 2 are
denoted as $M_\mathrm{d}(\beta =1)$ and
$M_\mathrm{d}(\beta =2)$,
respectively. Although the complex dependence of
quantities in equation~(\ref{eq:mdust}) makes it hard to
obtain an
analytical estimate of the uncertainty in $M_\mathrm{d}$,
we have confirmed that the uncertainty is around a factor
of 2 by varying the measured fluxes within the errors.
{}From Table \ref{tab:mdust}, we observe that
$M_\mathrm{d}(\beta =1)$ is systematically smaller than
$M_\mathrm{d}(\beta =2)$ by a factor of $\sim 2.5$,
mainly because of higher dust temperature for
$\beta =1$.

\begin{table*}
\centering
\begin{minipage}{150mm}
\caption{Some derived and compiled quantities.}\label{tab:mdust}
    \begin{tabular}{lccccccc}
      \hline
      Name     & $D\,^\mathrm{a}$ &
               $T_\mathrm{d}(\beta =1)\,^\mathrm{b}$ &
               $T_\mathrm{d}(\beta =2)\,^\mathrm{b}$ &
               $M_\mathrm{d}(\beta =1)\,^\mathrm{c}$ &
               $M_\mathrm{d}(\beta =2)\,^\mathrm{c}$ &
               $M_\mathrm{H\, I}\,^\mathrm{d}$ &
               $12+\log \mathrm{(O/H)}\,^\mathrm{e}$ \\
               & [Mpc] & [K] & [K] & [$M_\odot$] & [$M_\odot$] &
               [$M_\odot$] & \\ \hline
      UM 420   & 234 & ${97^{+249}_{-31}}$ & $50^{+20}_{-9}$ &
               $2.5\times 10^5$ & $8.7\times 10^5$ &
               --- & $7.93\pm 0.05$ \\
      Mrk 59   & 10.9 & $44^{+15}_{-7}$ & $32^{+7}_{-3}$ &
               $2.2\times 10^4$ & $6.2\times 10^4$ &
               --- & $7.99\pm 0.01$ \\
      Mrk 487  & 11.0 & $62^{+72}_{-16}$ & $40^{+16}_{-7}$ &
               $1.7\times 10^3$ & $5.0\times 10^3$ &
               $8.0\times 10^7$ & $8.06\pm 0.04$ \\
      SBS 1319+579 & 29.0 & $34^{+7}_{-4}$ & $26^{+4}_{-2}$ &
               $6.8\times 10^4$ & $1.8\times 10^5$ &
               $1.7\times 10^9$ & $8.10\pm 0.01$ \\ \hline
      II Zw 40 & 9.2 & $53^{+20}_{-9}$ & $36^{+8}_{-4}$ &
               $2.8\times 10^4$ & $8.3\times 10^4$ &
               $2.0\times 10^8$ & $8.15\pm 0.02$ \\
      Mrk 7    & 42.4 & $28^{+4}_{-2}$ & $23^{+3}_{-1}$ &
               $1.0\times 10^6$ & $2.7\times 10^6$ &
               $3.6\times 10^9$ & $8.54\pm 0.04$ \\
      Mrk 71   & 3.4 & $42^{+10}_{-5}$ & $31^{+5}_{-3}$ &
               $3.9\times 10^3$ & $1.1\times 10^4$ &
               $1.2\times 10^9$ & $7.83\pm 0.02$ \\
      UM 439   & 13.1 & $39^{+10}_{-5}$ & $30^{+4}_{-3}$ &
               $9.8\times 10^3$ & $2.7\times 10^4$ &
               $1.7\times 10^8$ & $7.98\pm 0.03$ \\
      UM 533   & 10.9 & $36^{+7}_{-5}$ & $28^{+4}_{-3}$ &
               $1.6\times 10^4$ & $4.2\times 10^4$ &
               $5.8\times 10^7$ & $8.10\pm 0.04$ \\
      II Zw 70 & 17.0 & $34^{+15}_{-5}$ & $27^{+8}_{-4}$ &
               $5.9\times 10^4$ & $1.5\times 10^5$ &
               $3.6\times 10^8$ & $8.11\pm 0.04$ \\
      II Zw 71 & 18.5 & $22^{+2}_{-2}$ & $19^{+1}_{-2}$ &
               $7.4\times 10^5$ & $1.8\times 10^6$ &
               $7.3\times 10^8$ & $8.24\pm 0.04$ \\
      \hline
    \end{tabular}

\medskip

 {
$^\mathrm{a}$ Distance estimated by using the Galactocentric
velocity taken fro NED by assuming a Hubble constant of
75 km s$^{-1}$ Mpc$^{-1}$ except for Mrk 71, whose distance
is taken from \citet{tolstoy95}. \\
$^\mathrm{b}$ Dust temperature estimated from the
140 $\mu$m--100 $\mu$m colour (Section \ref{subsec:dust_content}).
The emissivity index adopted is also indicated
($\beta =1$ or 2).\\
$^\mathrm{c}$ Dust mass estimated in
Section \ref{subsec:dust_content}. The emissivity index adopted
is also indicated ($\beta =1$ or 2).\\
$^\mathrm{d}$ H\,\textsc{i} gas mass estimated from
H\,\textsc{i} 21 cm line observations. For the latter seven
BCDs, the data are compiled in \cite{hirashita08}. The data for
Mrk 487 and SBS 1319+579 are obtained from \citet{hopkins02} and
\citet{huchtmeier07}, respectively.\\
$^\mathrm{e}$ Oxygen abundance compiled in \citet{hopkins02}
and \citet{hirashita02} (and the references therein). When
the error estimate is not available, we put a typical uncertainty in
the measurement (0.04) \citep{shi05}.
}
\end{minipage}
\end{table*}

As an indicator of dust abundance, we take dust-to-gas
ratio. For the gas mass, we adopt H\,\textsc{i} gas mass
$M_\mathrm{H\, I}$ estimated from H\,\textsc{i} 21 cm
emission observations. The data of $M_\mathrm{H\, I}$ are
already compiled by H08 for the common sample (i.e.\ the
latter seven BCDs in Table \ref{tab:mdust}). Among the
newly added sample, there is no available H\,\textsc{i} data
for UM 420 and Mrk 59, while the
data for Mrk 487 and SBS 1319+579 are obtained
from \citet*{hopkins02} and \citet{huchtmeier07}, respectively
(for the latter, we used the formula in
\citealt{lisenfeld98} to convert the H\,\textsc{i}
flux to $M_\mathrm{H\, I}$). The dust-to-gas ratio
$\mathcal{D}$ is defined as
$\mathcal{D}\equiv M_\mathrm{d}/M_\mathrm{H\, I}$.
We adopt $M_\mathrm{d}(\beta =1)$ for the dust mass,
but the following discussion does not change if
we adopt $M_\mathrm{d}(\beta =2)$.

\begin{figure*}
\begin{center}
\includegraphics[width=8.5cm]{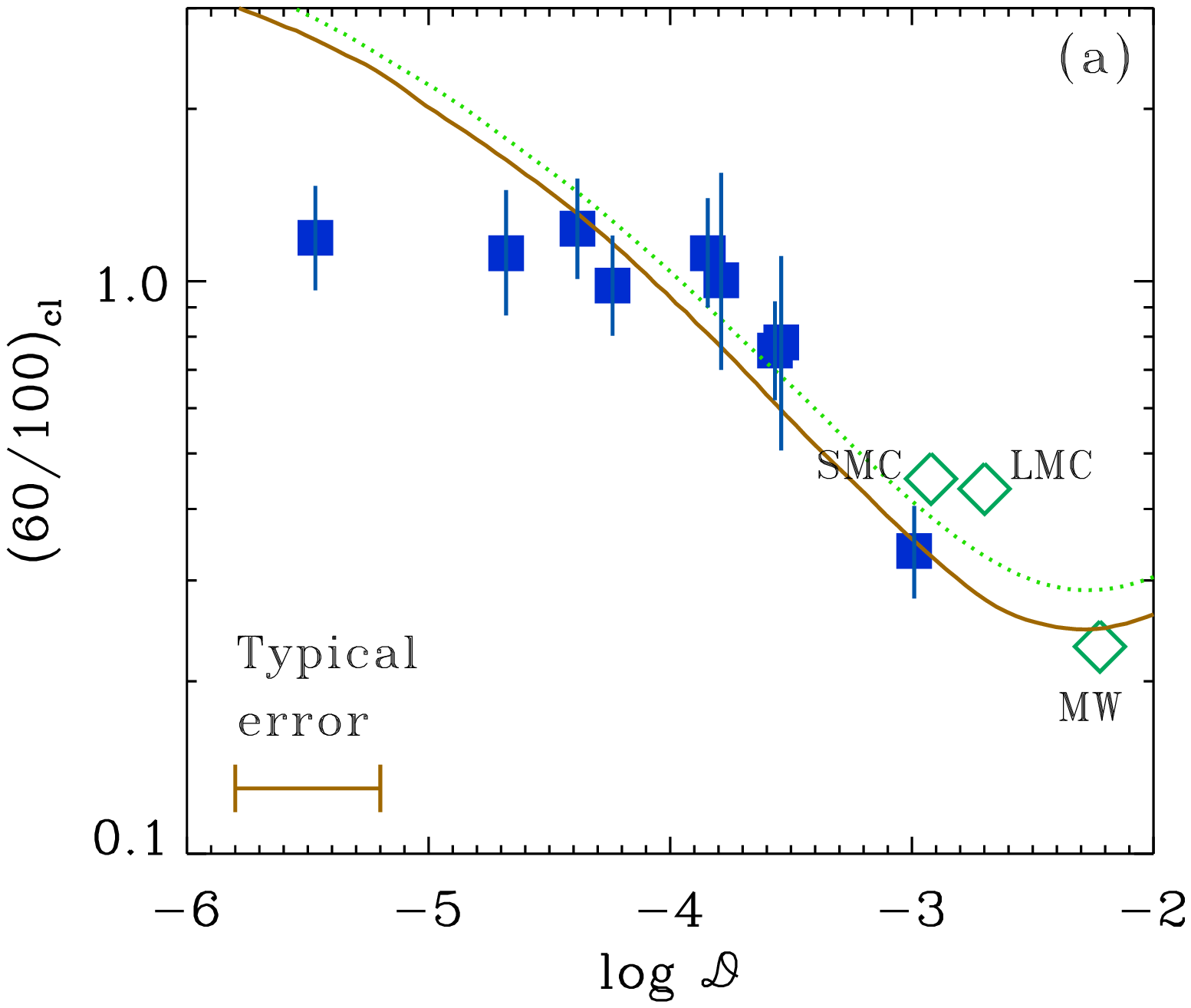}
\includegraphics[width=8.5cm]{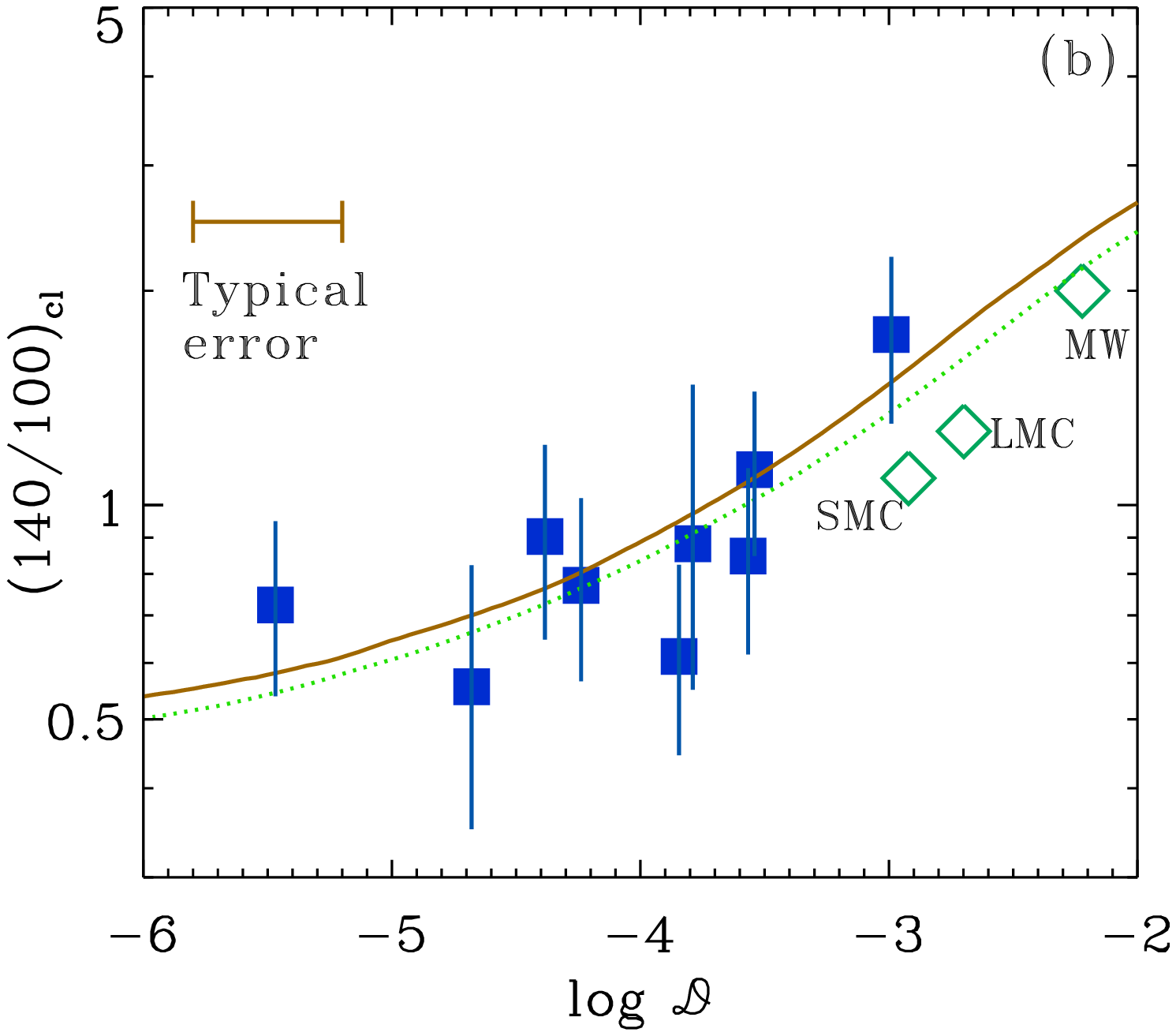}
\end{center}
\caption{Relation between far-infrared colour
($(60/100)_\mathrm{cl}$ and $(100/140)_\mathrm{cl}$ in
Panels a and b, respectively) and dust-to-gas
ratio. The filled squares with vertical error bars
represent the current BCD sample whose H \textsc{i} mass is
available. The typical error of dust-to-gas ratio which
comes from the error of estimated dust mass is shown by the
horizontal bar. The data points for the Milky Way,
the LMC and the SMC are also presented by the open diamonds.
The solid and dotted lines show the colour variation
expected from the model described by
equation (\ref{eq:chi_dg})  {with the optically thin
extreme ($A_V=0$)} for silicate and graphite,
respectively. The correlation coefficients for the BCD data
are $r=-0.65$ and $r=0.89$ for Panels a and b, respectively.
\label{fig:dg_clr}}
\end{figure*}

Fig.~\ref{fig:dg_clr} shows the relations between
FIR colour ($(60/100)_\mathrm{cl}$ or
$(140/100)_\mathrm{cl}$) and dust-to-gas ratio.
We observe correlations with correlation
coefficients $r=-0.65$ for the
$\log (60/100)_\mathrm{cl}$--$\log\mathcal{D}$
relation and $r=0.89$ for the
$\log (140/100)_\mathrm{cl}$--$\log\mathcal{D}$
relation. These correlations indicate that
the dust
temperature tends to be high in dust-poor objects.
We present not only the sample BCDs but also
the data of the Milky Way, the LMC and the SMC. The
dust-to-gas ratio of the Milky Way is assumed to be
0.006 \citep{spitzer78}, while the dust-to-gas ratios
of the LMC and the SMC are assumed to be
1/3 and 1/5 times of the value of the Milky Way
\citep{pei92}. For the FIR colours, we adopt the
peak of the contours in Fig.\ \ref{fig:clrdat}
for the Milky Way, and the average of the data points
for the LMC and the SMC. Those
three points also follow the trend found by the BCDs.

\begin{figure*}
\begin{center}
\includegraphics[width=8.5cm]{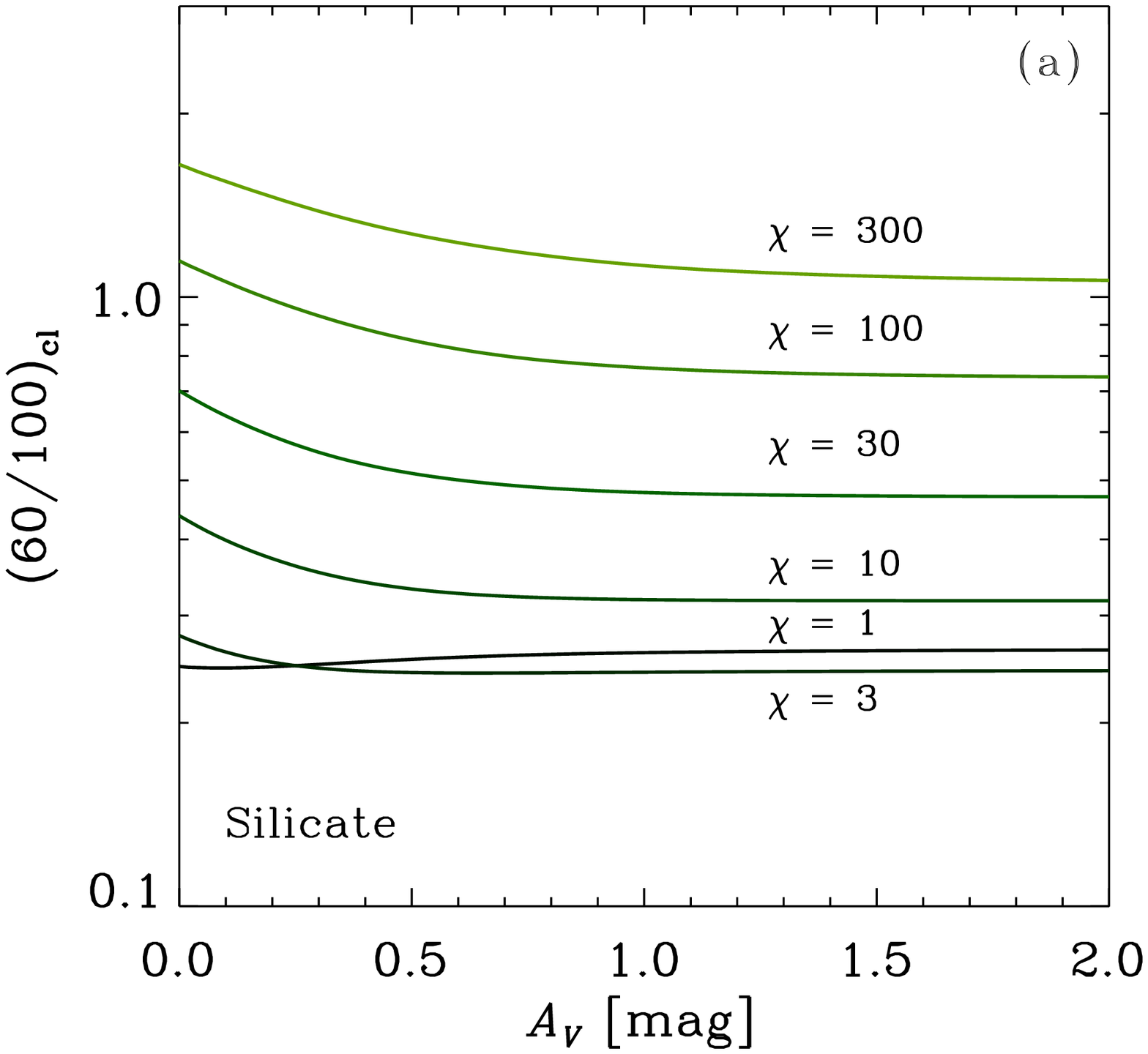}
\includegraphics[width=8.5cm]{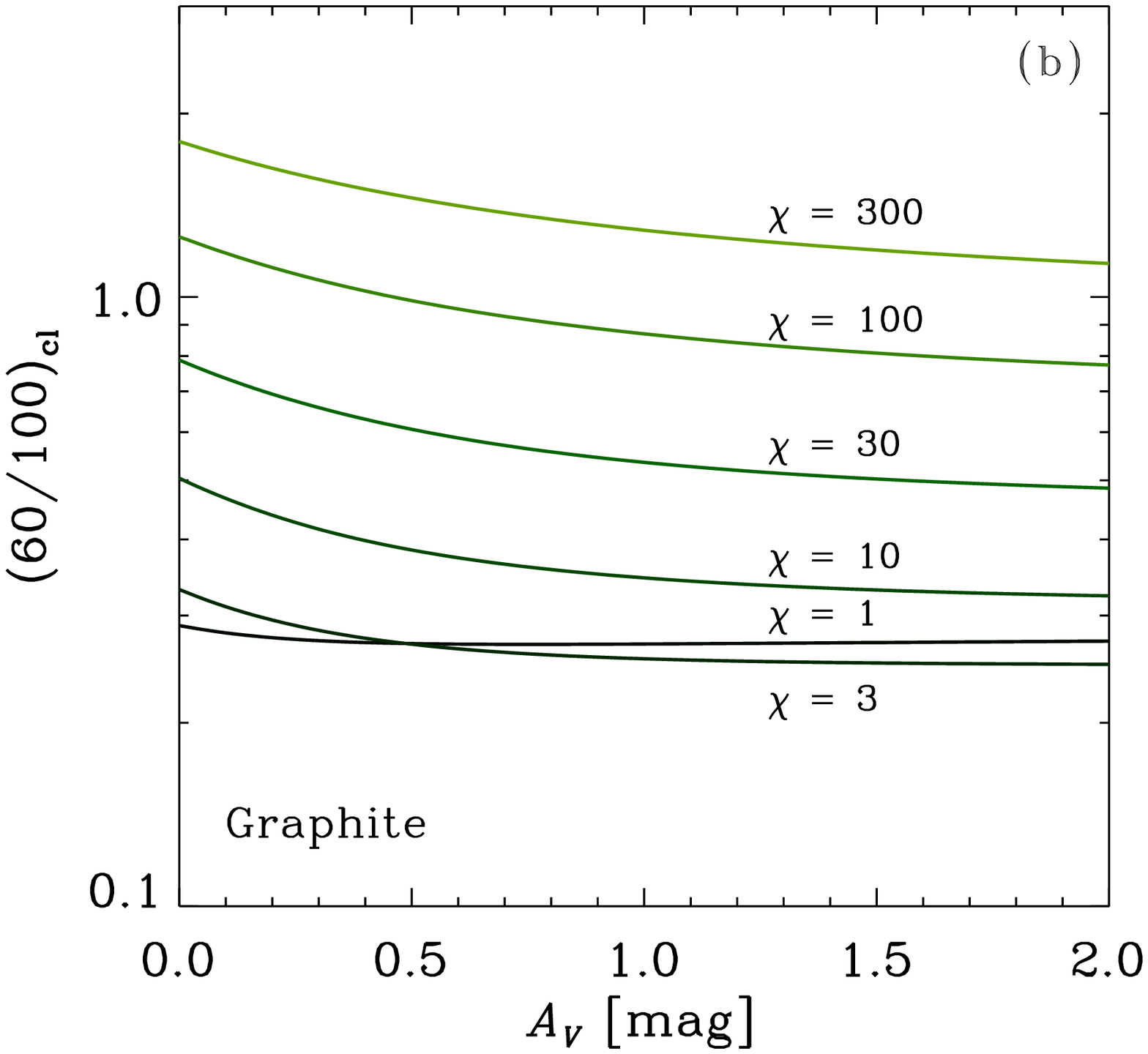}
\includegraphics[width=8.5cm]{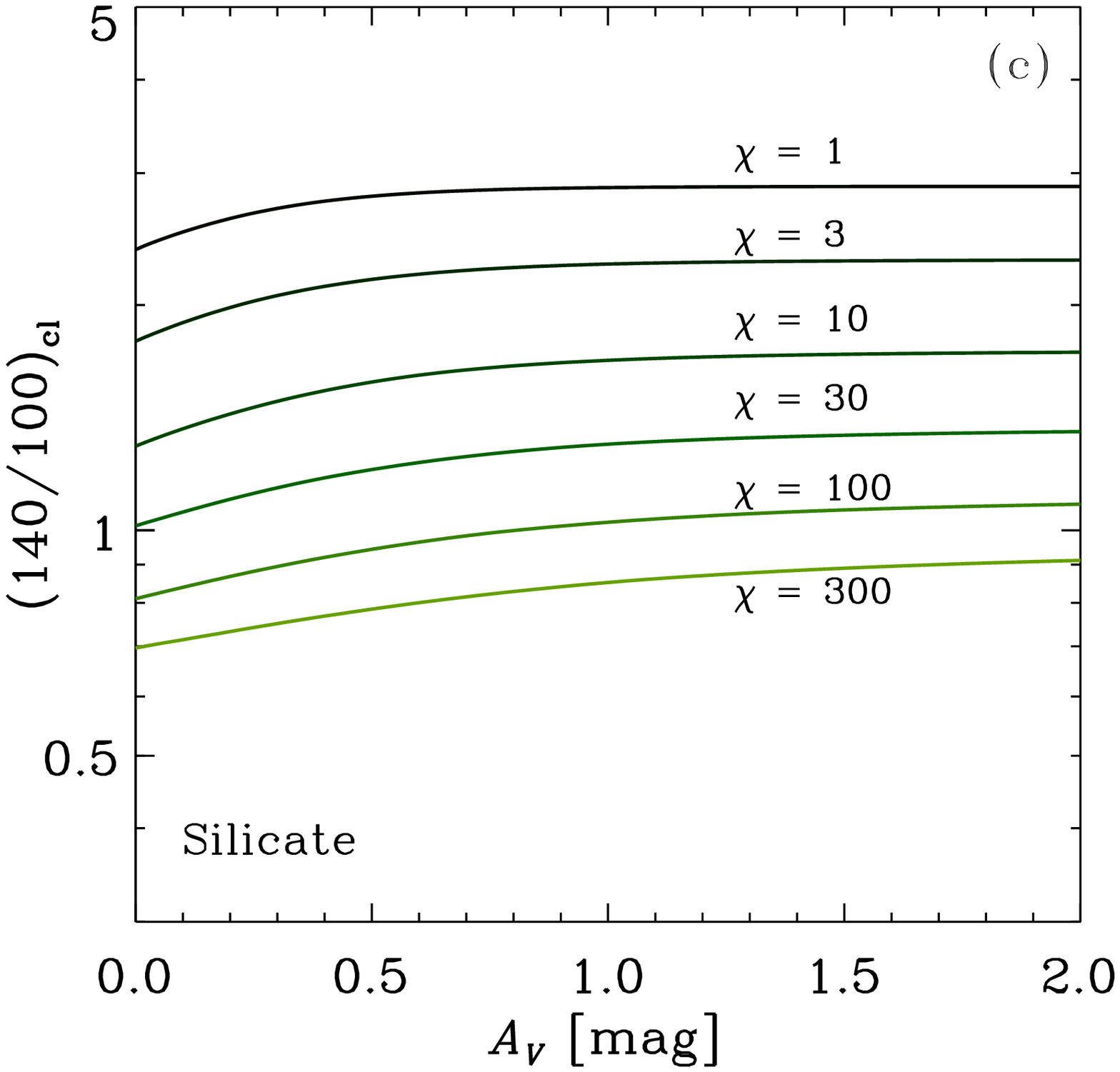}
\includegraphics[width=8.5cm]{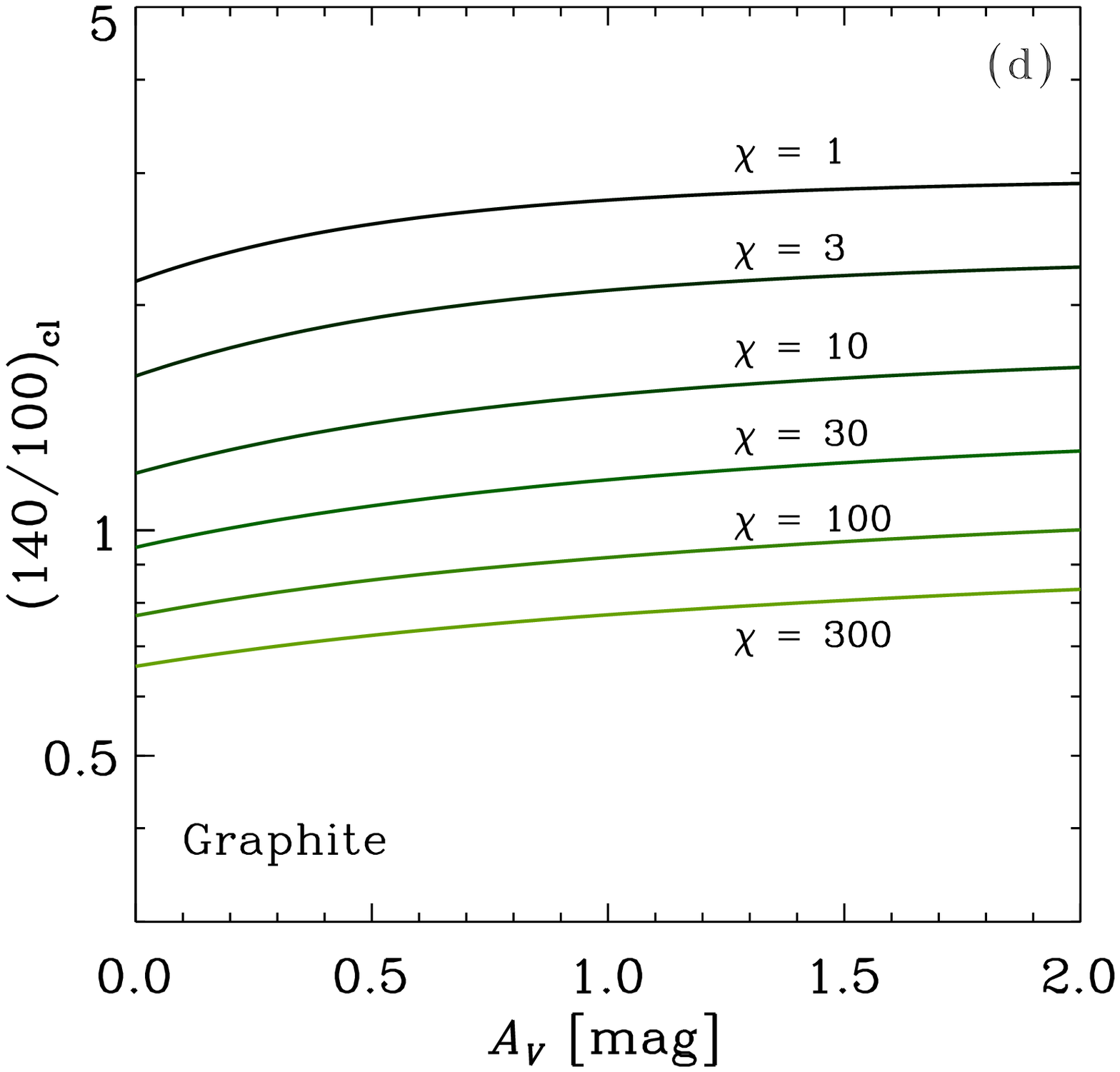}
\end{center}
\caption{Far-infrared colours ($(60/100)_\mathrm{cl}$ in
Panels a and b, and $(100/140)_\mathrm{cl}$ in Panels c
and d)  {for a foreground screen} as a function of
$A_V$. For the grain species we
examine silicate (Panels a and c) and graphite
(Panels b and d).
In each panel, different lines show different ISRF
intensities of $\chi =1$, 3, 10, 30,
100, and 300.
\label{fig:Av_clr}}
\end{figure*}

A possible interpretation of the correlation in
Fig.\ \ref{fig:dg_clr} is that the relatively low dust
temperatures in dust-rich BCDs result from the
shielding (extinction) of stellar radiation. In order to
examine this possibility, we calculate the relation
between FIR colours and $A_V$ by using the models
(same as Fig.\ \ref{fig:clr_varAv}). The results are shown
in Fig.\ \ref{fig:Av_clr}, where we find that both the
variation of $(60/100)_\mathrm{cl}$ and that of
$(140/100)_\mathrm{cl}$ with a single value of $\chi$
are too small to explain the observed large diversity in
these colours in Fig.\ \ref{fig:dg_clr}. The colours are not so
sensitive to the
change of $A_V$ partly because the strongest contribution
from $A_V\sim 0$, where the dust temperature is the highest,
is always present. In particular, if $A_V\ga 2$, the colour
is insensitive to $A_V$ since the dust temperature
at such deep optical depths is too low to contribute to
the total emission.
Thus, the variation of the FIR colours cannot
be reproduced only with a variation of dust optical depth,
but rather it should reflect the correlation (i) between
dust-to-gas ratio and dust heating itself
(i.e., ISRF), or (ii) between dust-to-gas ratio and dust
properties (i.e., absorption coefficient and/or
grain size distribution).

It is difficult to survey all the possible cases for (ii), since
there are few, if any, constraints on the dust properties
in BCDs. However, we can discuss some cases for (ii)
as follows. HHS07 show that $(140/100)_\mathrm{cl}$
changes only slightly along with the change of grain
size distribution, since it reflects the equilibrium grain
temperature. Thus, it is
hard to reproduce the clear correlation between
$(140/100)_\mathrm{cl}$ and dust-to-gas ratio only with
the change of grain size distribution. The change of
grain material as a function of dust-to-gas ratio may
change the grain absorption coefficient, leading to
variation of dust temperature as a function of dust-to-gas
ratio. However, we have shown that the colour--colour
relation of the BCDs can be explained by the
emissivity that also explains the colour--colour relation
of the Milky Way. In this context, it is not probable that
the dust property is the main driver for the change of
FIR colours.

The possibility (i) is interesting to investigate, since the
correlation in Fig.\ \ref{fig:dg_clr} is a natural extension
of the sequence of the Milky Way, the LMC, and the
SMC. For those ``nearby'' three galaxies, it is known
that the
typical ISRFs are different \citep[e.g.][]{welty06}, and
indeed the FIR colour
sequence can be explained by the variation of the ISRF
(\citealt{hibi06}, HHS07). Thus, in the following
subsection, we discuss a possibility that the
ISRF changes as a function of dust-to-gas ratio.

\subsection{Correlation between dust-to-gas ratio and
ISRF?}

Here we interpret the correlation between dust-to-gas
ratio and FIR colours as the correlation between
dust-to-gas ratio and ISRF. Since the main heating
source of dust grains is
UV photons coming from the young stellar
population \citep{buat96}, the dust-heating ISRF
reflects the
spatial concentration of recent star formation
activity. Thus, the correlation between dust-to-gas
ratio and ISRF indicates that
dust-poor galaxies tend to host concentrated
star-forming regions.

There are some theoretical models which suggest that
dust plays an active role in determining the properties
of star formation. Indeed \citet{hirashita04} show that
the shielding of UV heating photons plays an important
role in determining the strength of
star formation: If the optical depth of dust becomes
large, the UV heating photons are efficiently blocked
and the gas efficiently cools to produce a favourable
condition for the star formation. Thus, the optical
depth of dust in UV is important, and a burst of star
formation could become possible if the UV optical depth
exceeds one. In this scenario, the condition for
star formation can be written as
\begin{eqnarray}
\kappa\mathcal{D}\rho R\sim 1\, ,\label{eq:tau1}
\end{eqnarray}
where $\kappa$ is the UV absorption coefficient per
dust mass and $R$ is the size of the star-forming
region. On the other hand, we assume that a constant
fraction of gas mass is converted to stars:
$L\propto M_\mathrm{gas}$, where
$M_\mathrm{gas}(\sim\rho R^3)$ is the gas mass, and
$L$ is the stellar UV luminosity. Since
$\chi\propto L/R^2\propto M_\mathrm{gas}/R^2\sim
\rho R$,
we obtain $\chi\propto\mathcal{D}^{-1}$ by
using equation (\ref{eq:tau1}). Normalizing the
quantities to the Milky Way values (i.e.\ $\chi =1$
and $\mathcal{D}=0.006$), we obtain
\begin{eqnarray}
\chi =\left(\frac{\mathcal{D}}{0.006}\right)^{-1}\, .
\label{eq:chi_dg}
\end{eqnarray}

In Fig.\ \ref{fig:dg_clr}, we calculate the relation
between the FIR colours and $\mathcal{D}$ by assuming
equation (\ref{eq:chi_dg}).  {The optically thin
extreme (i.e.\ $A_V=0$) for the ISRF is adopted to focus
on the effect of $\chi$.} The
results indeed traces
the trend of the observational data. Thus, the above
argument on dust shielding of UV photons
(equation \ref{eq:tau1}) as the condition for
a burst of star formation is compatible with
the data.

Equation (\ref{eq:tau1}) implies that the size-density
relation of star-forming region becomes roughly
a relation with constant column density. Indeed, the
observational data presented by \citet{hirashita09}
support constant column density of star-forming regions
in BCDs (i.e.\ $\rho\propto R^{-1}$). \citet{pak98} also
find that the column density of gas is inversely
proportional to the dust-to-gas ratio based on
an analysis of photodissociation
regions in the Milky Way, the LMC, and the SMC, which
is consistent with our picture. \citet{mckee89}
show that an equilibrium between the gravitational
contraction and the energy injection from stars
is achieved at $A_V\sim 4$--8 based on the stability
argument on photoionization-regulated star formation, in
which shielding of photoionizing photons permits ambipolar
diffusion to proceed and stars to form. Although the
physical processes considered in \citet{mckee89} are
more detailed and sophisticated,
the necessity of shielding is common with our picture
expressed in equation (\ref{eq:tau1}).

\subsection{FIR colour and metallicity}
\label{subsec:metal}

The observational correlation between FIR colour and
dust-to-gas ratio shown in this paper has confirmed the
results of \citet{engelbracht08}, who show the correlation
between \textit{Spitzer} MIPS 70~$\mu$m--160~$\mu$m
colour and metallicity. Since there is a correlation between
dust-to-gas ratio and metallicity as shown by them and
also by H08, the correlation with metallicity is
equivalent to the correlation with dust-to-gas ratio.
For the detailed discussions on the relation between
dust-to-gas ratio and metallicity in BCDs, see H08
and \citet{lisenfeld98}.

\begin{figure*}
\begin{center}
\includegraphics[width=8.5cm]{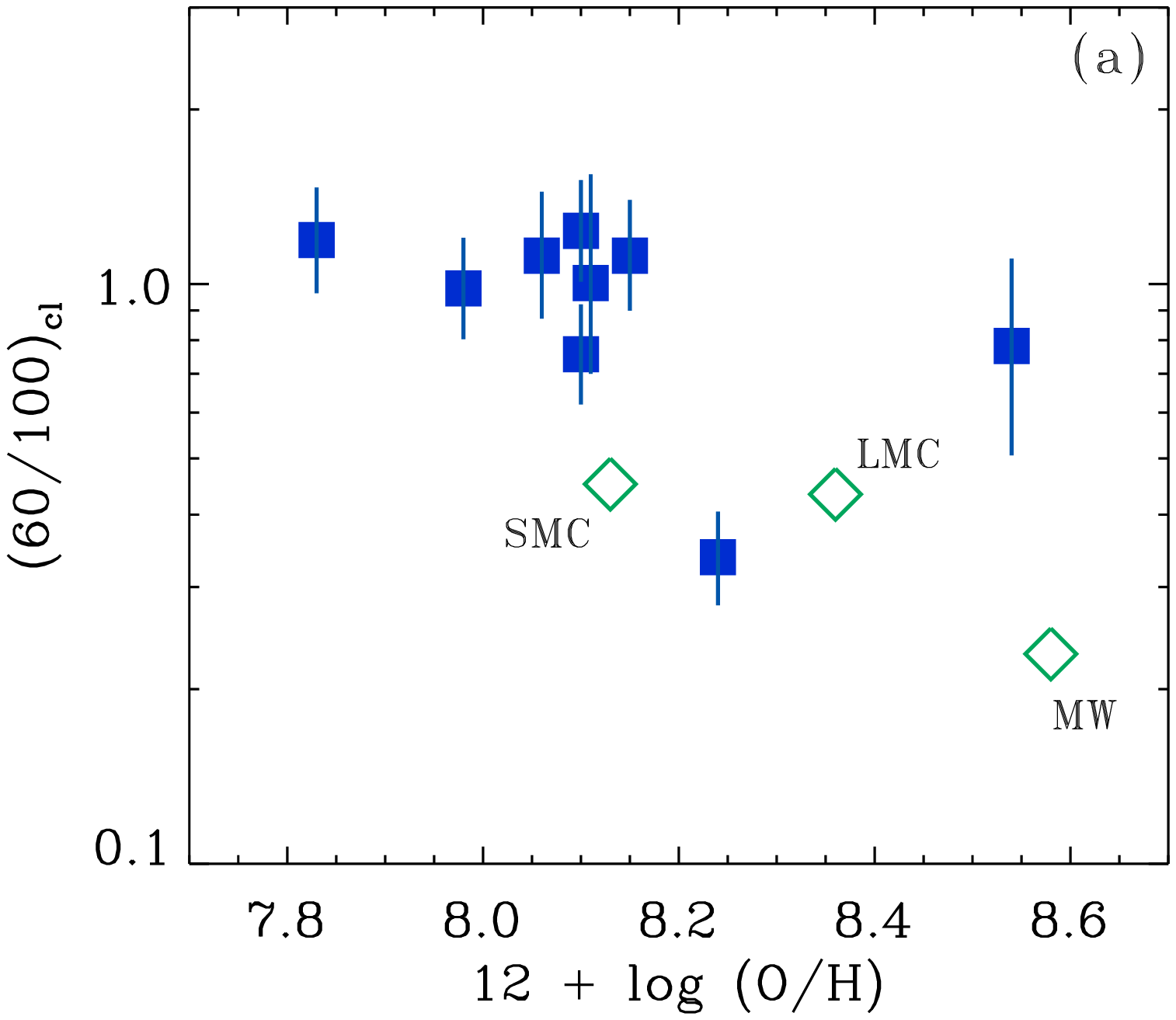}
\includegraphics[width=8.5cm]{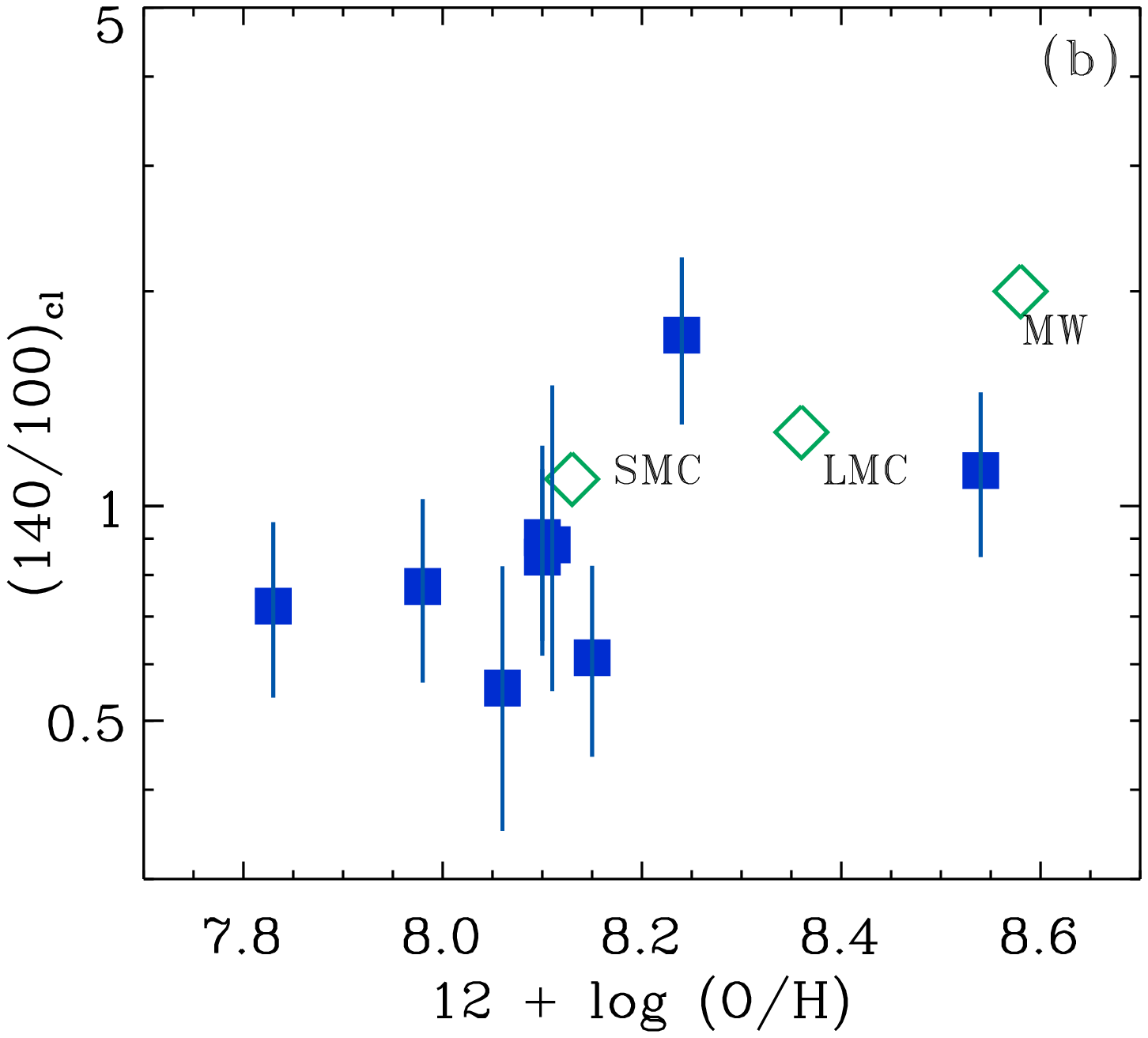}
\end{center}
\caption{Relation between FIR colour ($(60/100)_\mathrm{cl}$
and $(100/140)_\mathrm{cl}$ for Panels a and b, respectively)
and oxygen abundance. The filled squares with vertical error
bars represent the same BCD sample as in
Fig.\ \ref{fig:dg_clr}.
The data points for the Milky Way, the LMC and the SMC are
also presented by the open
diamonds.  {The uncertainty in the oxygen abundance
measurement is shown in Table \ref{tab:mdust}.}
The correlation coefficients for the BCD data are
$r=-0.45$ and $r=0.53$
for Panels a and b, respectively.
\label{fig:metal_clr}}
\end{figure*}

Fig.\ \ref{fig:metal_clr} shows the relation between FIR
colour and metallicity (oxygen abundance in gas phase) for
the BCD sample. The data of oxygen abundance are compiled
in \citet*{hirashita02} and \citet{hopkins02}, and summarized
in Table \ref{tab:mdust}. We observe correlations with
$r=-0.45$ for the relation between $\log (60/100)_\mathrm{cl}$
and $12+\log (\mathrm{O/H})$ and $r=0.53$ for the relation
between $\log (140/100)_\mathrm{cl}$ and
$12+\log (\mathrm{O/H})$.
We also show the data points for the Milky Way, the LMC,
and the SMC. The Milky Way gas metallicity is represented by
the Orion Nebula ($12+\log (\mathrm{O/H})=8.58$), which has
a similar metallicity to local B-type stars
\citep{peimbert87,gies92,kilian92,mathis00}. For the LMC
and the SMC, we adopt the oxygen abundance from
\citet{dufour84} ($12+\log (\mathrm{O/H})=8.43$ and 8.02,
respectively). We confirm that metal-poor galaxies tend to
have high dust temperatures.

We should keep in mind that the story of star formation
properties in terms of metallicity is not so simple
as we discussed above. \citet{engelbracht08} show that the
dust temperature shows the peak around
$12+\log (\mathrm{O/H})\sim 8$. Due to our small sample in
such a low metallicity range, we cannot confirm it, but
their results imply that the size and density of
star-forming region is not a monotonic function
of metallicity. There are also pieces of evidence that
there are two
categories of BCDs: ``active'' and ``passive''. The former
hosts compact and intense star formation activities, while
the latter has diffuse star formation properties
\citep{hunt02,hirashita04}. Even with a similar metallicity,
such a dichotomy exists, indicating that the star formation
properties are not determined simply by the metallicity.
Active BCDs tend to be luminous in FIR because of large
dust optical depth \citep{hirashita04}. Thus, it would
be possible that the current sample is biased to the
``active'' class. The presence of BCDs hosting cold dust
in Virgo Cluster \citep{popescu02} also implies that
such a bias could be present. This kind of bias should be
examined in
the future with more sensitive facilities.

Finally, it is worth questioning which of dust-to-gas
ratio and metallicity is more fundamental in regulating
dust temperature. Larger absolute values of correlation
coefficient found in Fig.~\ref{fig:dg_clr} than those in
Fig.~\ref{fig:metal_clr} imply that dust-to-gas ratio
rather than metallicity is more
strongly connected with dust temperature. Thus, we here
suggest that  dust-to-gas ratio is more fundamental than
metallicity in regulating the dust temperature. This should
be further examined with a larger sample.

\subsection{Importance of submillimetre data}
\label{subsec:submm}

It is expected that the radiative transfer effect, i.e.\
the shielding effect of dust can be seen more clearly in
submillimetre (submm) than in FIR, since the submm
emission can trace dust with lower temperature
\citep[e.g.][]{galliano03}. Although the submm data of
the current BCD sample are still lacking except for those
of II\,Zw\,40, future more sensitive
submm facilities such as \textit{Herschel} and ALMA
will increase the data.

\begin{figure}
\begin{center}
\includegraphics[width=8.5cm]{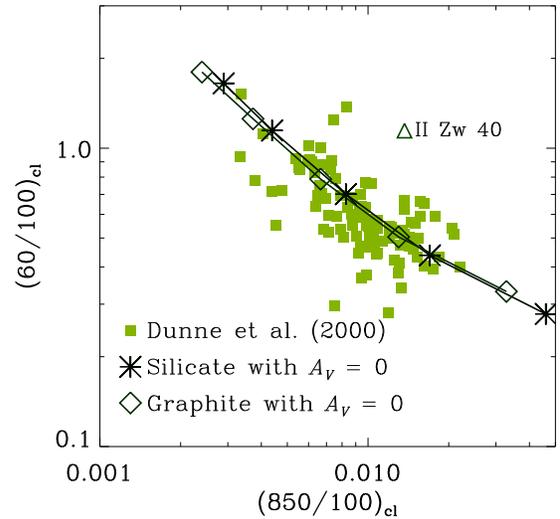}
\end{center}
\caption{$(60/100)_\mathrm{cl}$--$(850/100)_\mathrm{cl}$
relations for silicate and graphite (asterisks and open
diamonds, respectively). The different points correspond
to $\chi =3$, 10, 30, 100, and 300 from high to low
$(850/100)_\mathrm{cl}$.
 {The optically thin extreme for the ISRF ($A_V=0$) is
adopted.} The filled
squares indicate the observational data of nearby galaxies
in \citet{dunne00} and the open triangle shows the data
point of II Zw 40 taken from \citet{galliano05}.
\label{fig:clr_Av0_submm}}
\end{figure}

We adopt $\lambda =850~\mu$m as a representative submm
wavelength, and examine the relation between
$(850/100)_\mathrm{cl}$ and $(60/100)_\mathrm{cl}$;
that is, we adopt
$\lambda =850~\mu$m instead of
$\lambda =140~\mu$m, which is adopted in the rest of
this paper. In Fig.\ \ref{fig:clr_Av0_submm},
we present the colour--colour relation with
$A_V=0$ (i.e.\ without the radiative transfer effect;
 {Section \ref{subsec:colour}}).
We also show
the nearby galaxy sample observed at $\lambda =850~\mu$m
by \citet{dunne00}, who also list the \textit{IRAS}
60 $\mu$m and 100 $\mu$m fluxes. We observe that the
prediction traces the trend of the data, indicating
that the submm flux is naturally explained by the
natural extension of the FIR SED. This means that a
very cold component contributing only to the submm
flux is not necessary for a major part of the
nearby galaxies.

\begin{figure*}
\begin{center}
\includegraphics[width=8.5cm]{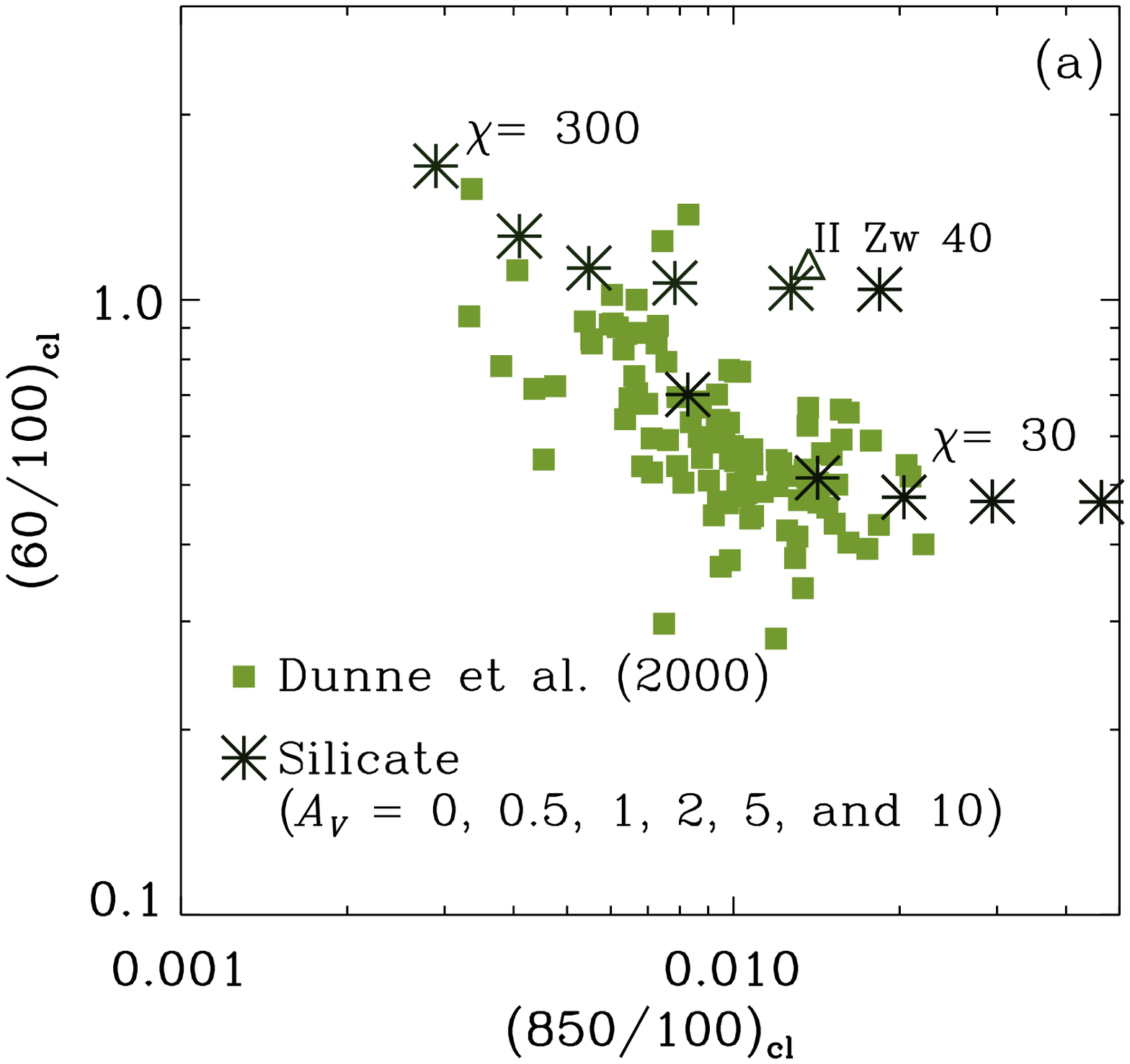}
\includegraphics[width=8.5cm]{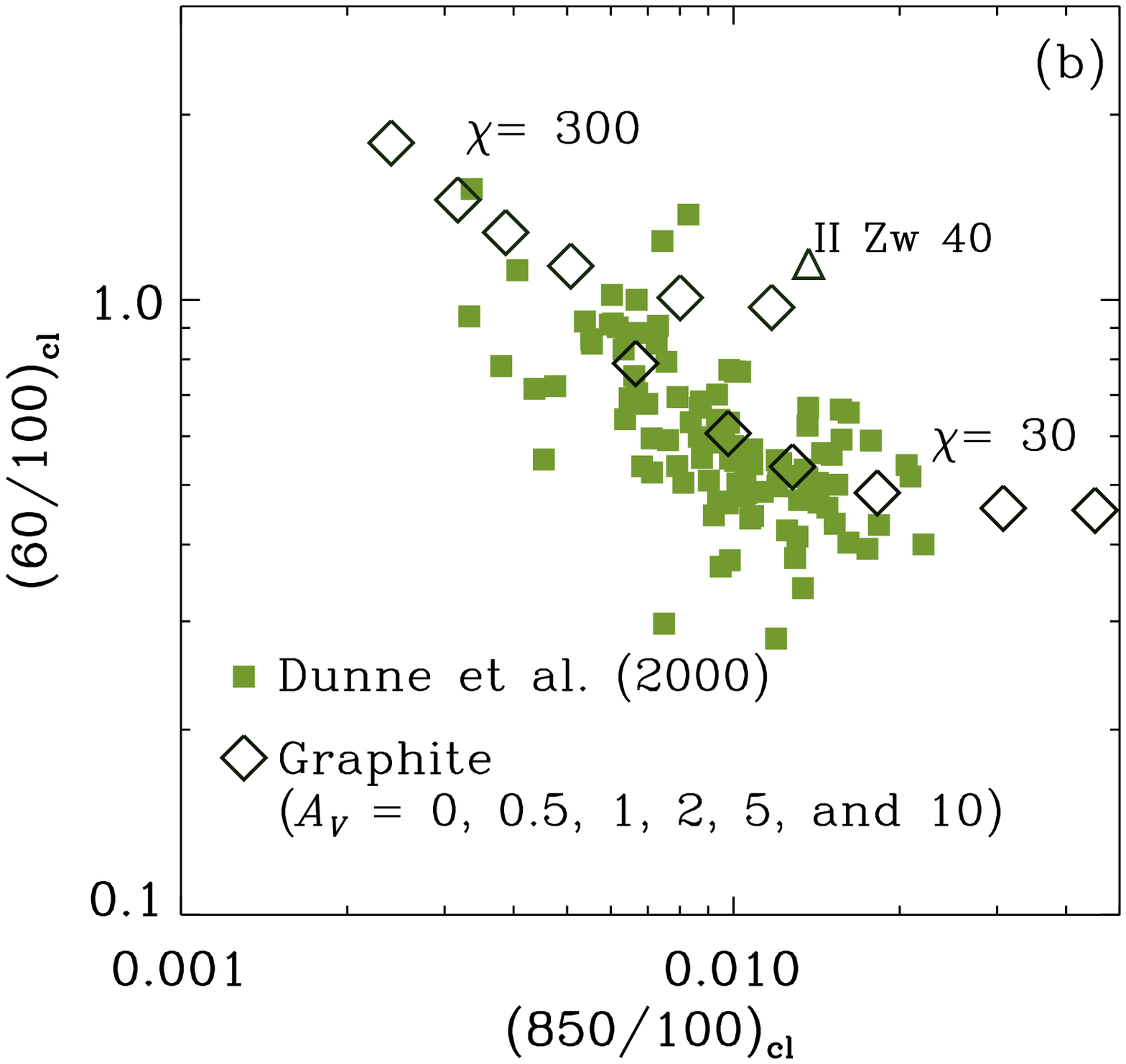}
\end{center}
\caption{$(60/100)_\mathrm{cl}$--$(850/100)_\mathrm{cl}$
relations for (a) silicate and (b) graphite (asterisks and open
diamonds, respectively).
The results with $\chi =30$ and 300 are shown. For each
value of $\chi$, $A_V=0$, 0.5, 1, 2, 5, and 10 correspond
to the data points from low to high $(850/100)_\mathrm{cl}$
(the point
for $\chi =30$ and $A_V=10$ is out of the range).
 {Note that $A_V=0$ is the optically thin extreme
for the ISRF.} The nearby
galaxy data taken from \citet{dunne00} are presented by the
filled squares.
\label{fig:clr_varAv_submm}}
\end{figure*}

\citet{galliano03,galliano05} argue that a very cold
component should be introduced for metal-poor dwarf
galaxies. In particular, II\,Zw\,40, which is
included also in the current sample, shows a clear
excess of the submm flux. In
Fig.\ \ref{fig:clr_Av0_submm}, we also plot the data
point taken from \citet{galliano05} for II\,Zw\,40
(see also \citealt{hunt05}). Indeed, we observe a
clear deviation from the model prediction, which can
be interpreted as a significant contribution from a very
cold component. They also propose that such very cold
dust should be located in an environment where the
stellar radiation is strongly shielded. In order to
examine the effect of shielding (i.e.\
radiative transfer), we show the results with
$A_V>0$ in Fig.\ \ref{fig:clr_varAv_submm}.
We observe that $(850/100)_\mathrm{cl}$ increases
even for $A_V>2$
while $(60/100)_\mathrm{cl}$ is almost constant for
such large
$A_V$. This confirms that 850 $\mu$m flux is also
sensitive to the shielded cold dust grains.
Fig.\ \ref{fig:clr_varAv_submm} also demonstrates
that the data point of II\,Zw\,40 is
explained with $A_V\sim 5$ with
$\chi\sim 300$. Thus we conclude the existence
of shielded cold dust in II Zw 40
(and in other galaxies whose data points deviate
rightwards in Fig.\ \ref{fig:clr_Av0_submm}).
\citet{galliano05} also show that the UV ISRF of
II Zw 40 is higher than that of the Milky Way by two
orders of magnitude.

Although our interpretation of the submm excess
as a contribution from the shielded cold dust
component is in agreement with the analysis by
\citet{galliano05}, there is
quantitative difference between our results and
theirs. \citet{galliano05} adopted
$Q_\mathrm{abs}\propto\lambda^{-2}$ at
$\lambda >100~\mu$m. If the absorption efficiency is
normalized to the value at $\lambda =100~\mu$m,
the absorption efficiency adopted in this paper
(equation \ref{eq:reach}) is 2 times larger than that
used in \citet{galliano05} at $\lambda =850~\mu$m.
Therefore, we require less very cold dust component
than \citet{galliano05} by a factor of 2.

\subsection{Common dust optical properties?}

In this paper, we have adopted a common FIR dust optical
properties; that is, we have applied the same absorption
efficiency $Q_\mathrm{abs}$ and grain size distribution
as adopted in our previous model for the Milky Way and
the Magellanic Clouds (HHS07). The same dust optical
properties are also consistent with the FIR colours of
BCDs. \cite{hibi06} and HHS07 also argue that those dust
optical properties also provide a good statistical fit to
the FIR colours of nearby galaxies. Therefore, we suggest
that the absorption efficiency in the form of
equation (\ref{eq:reach}), which
\citet{reach95} adopted to fit the dust emission spectra
in the Milky Way, is generally
applicable in galactic environments. As shown above,
the submm emission from nearby galaxies can also be
explained by the same dust emissivity.

It is also important to stress that the classical
dust emissivity model by \citet{draine84} cannot
explain the FIR colour--colour relation of the nearby
galaxies (\citealt{hibi06}; HHS07). As indicated by
the emissivity assumed by \citet{reach95}, it is
better to adopt $\beta\sim 1$ for
$100~\mu\mathrm{m}\la\lambda\la\lambda_1\sim 200~\mu$m
rather than to assume $\beta\sim 2$ in the entire
FIR wavelength range longer than 100~$\mu$m. Some
amorphous materials indeed show $\beta <2$
\citep[e.g.][]{agladze96}. Amorphous grains are
expected to form by the irradiation of cosmic
rays \citep{jager03}.

Theoretically the break of the dust emissivity at
$\lambda_1$ might be associated with the energy
splitting of the ground state because of irregular
amorphous structure.
\citet{meny07} show that the absorption coefficient
of amorphous dust has a break at
$\lambda\sim\lambda_\mathrm{m}$, which corresponds to
the cut-off energy $\hbar\omega_\mathrm{m}$ of the
energy splitting due to the amorphous structure
($\omega_\mathrm{m}\equiv 2\pi c/\lambda_\mathrm{m}$).
Therefore, a possible interpretation is that
$\lambda_\mathrm{m}\sim\lambda_1$.
Since \citet{meny07} suggest
$\lambda_\mathrm{m}\sim 700~\mu$m based on
experimental data, the cut-off energy corresponding
to $\lambda_\mathrm{m}\sim 200~\mu$m may be too
high. Nevertheless it is still worth investigating
amorphous materials with higher cut-off energy as
a candidate of cosmic dust.

\section{Conclusion}\label{sec:conclusion}

We have investigated the properties of FIR emission
of a sample of BCDs observed by \textit{AKARI},
especially focusing on FIR colours and dust temperature.
We have
utilized the data at $\lambda =65~\mu$m, 90~$\mu$m, and
140~$\mu$m, and have examined the relation between
60~$\mu$m--100~$\mu$m colour,  $(60/100)_\mathrm{cl}$,
and 140~$\mu$m--100~$\mu$m colour
$(140/100)_\mathrm{cl}$. Then, we have
found that the FIR colours of the BCDs are located at a
natural high-temperature extension of the DIRBE data of
the Milky Way, the LMC and the SMC on the colour--colour
diagram. We have explained the FIR colours also
theoretically by assuming the same absorption efficiency,
which may be appropriate for amorphous dust grains,
and the same grain size distribution as the Milky Way
dust. We have also shown that it is not easy to
distinguish between a large dust optical depth and a
low dust temperature only with FIR colours although
addition of submillimetre data relax this degeneracy.

In order to examine if the dust optical depth plays an
important role in determining the dust temperature, we
have investigated the correlation between FIR colour (dust
temperature) and dust-to-gas ratio. We have found that
the dust temperature tends to become high as the
dust-to-gas ratio decreases
as would be expected from the shielding effect of
stellar radiation by dust.
However, we fail to explain this trend quantitatively
only by the effect of
dust optical depth. Thus, we conclude that there is
a relation between dust-to-gas ratio and interstellar
radiation field (ISRF) intensity.
This relation is consistent with a ``constant dust
optical depth'' of star-forming regions in BCDs,
implying that dust extinction
plays an important role in determining the condition
for a burst of star formation.

The correlation between dust-to-gas ratio and dust
temperature is equivalent to that between metallicity and
dust temperature
found in \citet{engelbracht08}, since there is a correlation
between dust-to-gas ratio and metallicity. By comparing
the correlation strengths, we propose that dust-to-gas ratio
is more fundamental than metallicity in regulating dust
temperature, although we will have to confirm this with
a larger sample.

\section*{Acknowledgments}
We thank the anonymous referee for useful comments which
improved this paper considerably.
We are grateful to T. Onaka, H. Kaneda, and Y. Hibi for helpful
discussions. We thank all members of \textit{AKARI} project for
their continuous help and support. This research has made
use of the NASA/IPAC Extragalactic Database (NED), which
is operated by the Jet Propulsion Laboratory, California
Institute of Technology, under contract with the National
Aeronautics and Space Administration.

\end{document}